\documentclass[aps,prd,twocolumn,nofootinbib,showpacs,superscriptaddress]{revtex4-1}

\usepackage{multirow, makecell}
\usepackage{amsfonts}
\usepackage{amsmath}
\usepackage{amssymb}
\usepackage{bm}
\usepackage{dcolumn}
\usepackage{epsfig}
\usepackage{graphicx}
\usepackage{graphics}
\usepackage[latin1]{inputenc}
\usepackage{latexsym}
\usepackage{rotating}
\usepackage[colorlinks=true]{hyperref}
\usepackage[usenames]{color}
\usepackage{yfonts}
\usepackage{xspace} 
\usepackage{mathrsfs}
\usepackage{subfigure}
\usepackage{enumitem}
\usepackage{tabularx}
\usepackage{booktabs}
\usepackage{siunitx}
\usepackage{array}
\usepackage{placeins}
\usepackage[normalem]{ulem}
\newcolumntype{L}[1]{>{\raggedright\let\newline\\\arraybackslash\hspace{0pt}}m{#1}}
\newcolumntype{C}[1]{>{\centering\let\newline\\\arraybackslash\hspace{0pt}}m{#1}}
\newcolumntype{R}[1]{>{\raggedleft\let\newline\\\arraybackslash\hspace{0pt}}m{#1}}

\usepackage{ulem}
\newcommand{\be}{\begin{equation}}
\newcommand{\ee}{\end{equation}}
\newcommand\ba{\begin{eqnarray}}
\newcommand\bse{\begin{subequations}}
\newcommand\ea{\end{eqnarray}}
\newcommand\ese{\end{subequations}}

\newcommand{\nn}{\nonumber}

\newcommand{\BD}{{\mbox{\tiny BD}}}
\newcommand{\GR}{{\mbox{\tiny GR}}}
\newcommand{\GRtwo}{{\mbox{\tiny GR2}}}
\newcommand{\GRfour}{{\mbox{\tiny GR4}}}
\newcommand{\GRsix}{{\mbox{\tiny GR6}}}
\newcommand{\GReight}{{\mbox{\tiny GR8}}}

\allowdisplaybreaks
\begin{document}

\title{Eccentricity Matters: \\ Improvement of Constraints on Jordan-Brans-Dicke-Fierz Theory \\ with Gravitational Waves from Eccentric Compact Binary Inspirals}

\title{Improved Constraints on Modified Gravity with Eccentric Gravitational Waves}

\author{Sizheng Ma}
\email{sma@caltech.edu}
\affiliation{TAPIR, Walter Burke Institute for Theoretical Physics, California Institute of Technology, Pasadena, California 91125, USA}
\affiliation{Department of Physics and Center for Astrophysics, Tsinghua University, Haidian District, Beijing 100084, China}
\author{Nicol\'as Yunes}
\email{nyunes@illinois.edu}
\affiliation{eXtreme Gravity Institute, Department of Physics, Montana State University, Bozeman, Montana 59717, USA}
\affiliation{Department of Physics, University of Illinois at Urbana-Champaign, Urbana, IL 61801, USA}

\date{\today}

\begin{abstract}
Recent gravitational wave observations have allowed stringent new constraints on modifications to General Relativity (GR) in the extreme gravity regime. 
Although these observations were consistent with compact binaries with no orbital eccentricity, gravitational waves emitted in mildly eccentric binaries may be observed once detectors reach their design sensitivity. 
In this paper, we study the effect of eccentricity in gravitational wave constraints of modified gravity, focusing on Jordan-Brans-Dicke-Fierz theory as an example. 
Using the stationary phase approximation and the post-circular approximation (an expansion in small eccentricity), we first construct an analytical expression for frequency-domain gravitational waveforms produced by inspiraling compact binaries with small eccentricity in this theory. 
We then calculate the overlap between our approximate analytical waveforms and an eccentric numerical model (TaylorT4) to determine the regime of validity (in eccentricity) of the former. 
With this at hand, we carry out a Fisher analysis to determine the accuracy to which Jordan-Brans-Dicke-Fierz theory could be constrained given future eccentric detections consistent with General Relativity.  
We find that the constraint on the theory initially deteriorates (due to covariances between the eccentricity and the Brans-Dicke coupling parameter), but then it begins to recover, once the eccentricity is larger than approximately $0.03$. 
We also find that third-generation ground-based detectors and space-based detectors could allow for constraints that are up to an order of magnitude more stringent than current Solar System bounds.
Our results suggest that waveforms in modified gravity for systems with moderate eccentricity should be developed to maximize the theoretical physics that can be extracted in the future.  
\end{abstract}

\maketitle

\section{Introduction}

The detection of gravitational waves (GWs) from merging black hole binaries \citep{gw150914,gw151226,gw170104,gw170608,gw170814} and neutron stars \citep{gw170817} has started a new era in astrophysics. Those signals were consistent with black holes moving in quasi-circular orbits \citep{PhysRevLett.116.241102,2041-8205-818-2-L22}, a result consistent with General Relativity's prediction that binaries circularize via GW emission~\citep{Peters63,Peters64,PhysRevLett.116.221101}. However, recent studies \citep{Fabio12,Johan14,Todd11,East13} show that several different astrophysical mechanisms could lead to inspiral signals that enter the sensitivity band of GW detectors with non-negligible eccentricity. An example of these are three-body interactions in hierarchical triples that live in galactic nuclei and globular clusters; the Kozai-Lidov mechanism may be significant in such systems, and this can drive oscillations in the eccentricity of the inner binary. Another example is the segregation of stellar-mass black holes toward galactic nuclei that harbor a supermassive black holes; this may cause high eccentricity encounters that form binaries with some eccentricity in the LIGO band \citep{OLeary09}. A third example consists of eccentric double white dwarf binaries formed in globular clusters, which are expected to be detectable by LISA \citep{Willems08}. A final example is the evolution of supermassive BH (SMBH) binaries in galactic nuclei, which can lead to orbits with eccentricities around $0.05$--$0.2$ when the low harmonics of the GW enter the LISA band \citep{Berentzen09}. 

Even if eccentric binaries are not detectable in the current observing runs of advanced LIGO and Virgo, eccentric binaries will be detected by both second- and third-generation detectors once they reach their design sensitivity, as argued by multiple authors (see e.g.~\citep{Huerta14} and references therein). Reference~\cite{Huerta14} found that advanced LIGO-type detectors could detect approximately $0.1$--$10$ events per year out to redshifts $z\sim0.2$, while an array of Einstein Telescope (ET) detectors could detect hundreds of events per year to redshift $z\sim2.3$. According to~\citep{LIGOwhite}, advanced LIGO (aLIGO) will be upgraded to A+ by 2022 and to Voyager by 2027, although these dates are likely to slip somewhat. Third-generation detectors, like ET and Cosmic Explorer (CE), are also planned in the 2030s. The space-based gravitational wave detector, LISA, is expected to be launched in the mid 2030s \citep{Amaro-Seoane17}. Given these plans for improved GW detectors, the accurate and efficient inclusion of eccentricity in GW models is both interesting and timely. 

One could in principle use quasi-circular waveform models to detect inspiraling eccentric binaries, but this would be inefficient and dangerous. Inappropriate waveforms can lead to either a significant loss of signal-to-noise ratio (SNR) \citep{Martel99,Brown10,Porter10,Tessmer08,Huerta:2013qb} or a systematic bias in parameter estimation \citep{Cutler07,Sun15}, which could then lead to incorrect astrophysical inferences. For example, Refs.~\cite{Brown10} and~\cite{Huerta:2013qb} showed that eccentric waveform models are needed to detect BH and NS binaries with eccentricities larger than $0.1$ and $0.4$ respectively. But even if the signal is detected, Ref.~\cite{Sun15} showed that systematic errors would be introduced in the recovered parameters that would dominate over statistical ones at SNRs larger than 10.  

For this reason, the effort to construct eccentric waveform models has ramped up over the last decade. The first studies of eccentric waveforms started perhaps with the seminal work by Peters and Mathews \citep{Peters63,Peters64}, who computed the energy and angular momentum flux from eccentric binary inspirals. The GW polarization states for eccentric inspirals were first presented by Wahlquist in the late 1980s~\citep{Wahlquist87} to leading order in the post-Newtonian (PN) expansion\footnote{The PN approximation is an expansion in weak-fields and small velocities, quantified by the ratio of the orbital velocity to the speed of light. Terms of $N$PN order are suppressed by factors of ${\cal{O}}(v^{2N}/c^{2N})$ relative to the leading-order term~\cite{Blanchet:2002av}.}. This model was extended to 1PN order in \citep{Junker92}, 1.5PN order in \citep{Blanchet93}, 2PN order in \citep{Gopakumar02} and elements of the 3PN calculations were computed in \citep{Damour04,Arun08,Arun082,Arun09}.

Although the ingredients to compute eccentric waveforms existed, more work had to be carried out to cast the model in a form suitable for data analysis studies, which operate in the frequency domain. The eccentric contributions to the Fourier phase of eccentric waveforms were first studied in Ref.~\citep{Krolak95} using the stationary phase approximation (SPA), a small eccentricity expansion valid to $\mathcal{O}(e_0^2)$, and to leading Newtonian order in the PN approximation. These waveforms were then extended to 2PN order in~\citep{Favata14} and 3PN order in~\citep{Moore16}. \citeauthor{Yunes09} \citep{Yunes09} proposed a formal double expansion in small eccentricity and small velocities, the post-circular (PC) approximation, to extend analytical quasi-circular waveforms (in the time- and frequency-domains) to eccentric ones. As a proof-of-principle, they computed Fourier waveforms in the SPA to leading Newtonian order in the PN approximation but to $\mathcal{O}(e_0^8)$. Based on this work, several efforts have been carried out since then to generalize this result to higher PN orders~\citep{Tessmer10,Tessmer102,Huerta14,Tanay16}; among these, \citeauthor{Tanay16} extended the PC approximation to 2PN and $\mathcal{O}(e_0^6)$. Recently, there has been work to create waveform models valid beyond the post-circular approximation, but we will not study those here~\cite{Moore:2018kvz,Moore:2019xkm}.

The analytic waveform models described above have allowed for parameter estimation studies of the effect of eccentricity. \citeauthor{Sun15} \citep{Sun15} used a high-PN order, PC model to show through a Fisher study that the accuracy of parameter recovery is enhanced by eccentricity in the signal. \citeauthor{Ma17} \citep{Ma17} further found that the angular resolution of a network of ground-based detectors can be improved by factor of $1.3\sim2$ due to eccentricity. In Ref. \citep{Mikoczi12}, \citeauthor{Mikoczi12} found that the precision of source localization for SMBHs detected by LISA improves significantly as a result of eccentricity. 

Given these results, one expects that eccentricity should improve the ability of detectors to constrain modified gravity theories, one of the primary science-drivers of ground- and space-based detectors~\cite{Yunes:2013dva}. In order to study this concretely, we focus on a particular example, scalar-tensor (ST) gravity, and in particular, on its simplest incarnation: Jordan-Brans-Dicke-Fierz theory \citep{Brans61}. This theory adds a scalar field that couples directly to the metric tensor, thus introducing modifications to Solar System observables and to the strong equivalence principle~\citep{Will:2010gnb,Eardley75,Will77,Will+Zag89}. The strength of the deviations are controlled by a (constant) coupling parameter, $\omega$, with the theory reducing to Einstein's when $\omega\to\infty$. The most stringent constraint, $\omega> 40,000$, comes from observations of the Shapiro time-delay through tracking of the Cassini probe~\citep{Bertotti03}. Although this theory is already stringently constrained, it serves as a good training ground to develop eccentric waveforms in modified gravity and to study the effect of eccentricity in possible constraints. 

GW observations of mixed BH-NS binaries should allow for independent constraints on ST theory through tests of the strong equivalence principle. Will \citep{Will94} was the first to derive the corrections to the Fourier phase of quasi-circular GWs to leading Newtonian order. Through a Fisher analysis, he found that future GW observations of mixed binaries could bound $\omega > 10^{3}$ with aLIGO. Later studies showed that much more stringent constraints, of the order of $\omega > 10^{5}$, could be achieved with GW observations of extreme mass-ratio inspirals with LISA~\citep{Scharre02,Will04,Yunes:2011aa}. The effect of spin was investigated in~\citep{Berti05} and shown to deteriorate the bound, while the effect of eccentricity and precession was included in the GR sector only in~\cite{Yagi10} and shown to improve the constraint. 

In this paper we carry out a systematic study of the effect of eccentricity in projected constraints on Jordan-Brans-Dicke-Fierz theory with both ground- and space-based detectors. We first calculate the ST corrections to the temporal and frequency evolution of the eccentricity during the inspiral to $\mathcal{O}(1/\omega)$. With this at hand, we then construct an analytic, frequency-domain waveform model in this theory for eccentric, inspiraling binaries in the PC approximation. The GR sector is modeled to 3PN order, including all eccentric corrections known at each PN order. The ST sector is here calculated for the first time to ${\cal{O}}(e_{0}^{8})$ and to leading Newtonian order in the PN approximation\footnote{Higher PN order corrections can be introduced in the future, once these are calculated; this calculation, however, goes well beyond the scope of this paper.}. We find that the eccentric ST corrections, just like in the GR case, introduce \emph{negative} PN order corrections, relative to the leading Newtonian order term in the quasi-circular limit. Such terms are very large at large separations (or small velocities) provided the eccentricity is not vanishingly small, thus enhancing the importance of ST terms in the GW phase evolution and possibly allowing for more stringent constraints given signals consistent with GR.     

We then carry out an overlap analysis to determine the regime of validity in eccentricity of our analytic ST model because it relies on the PC approximation. To do so, we focus on GR and first construct a purely numerical inspiral model in the time-domain, which we then discrete Fourier transform into the frequency-domain. Such a numerical model is similar to the TaylorT4 model~\cite{Buonanno09}, but for eccentric waveforms in GR, as already discussed e.g.~in~\cite{Moore:2018kvz}. We then derive new analytic expressions to rapidly maximize the overlap over the phase offset when there are multiple harmonics present in the waveforms, provided one of them is dominant; this result is similar to that presented in~\cite{Moore:2018kvz}. Next we calculate the match, i.e.~the overlap maximized only over time and phase offset but not over system parameters, between our analytic model and the numerical one as a function of initial eccentricity. Demanding that the match is larger than 97\% provides a (minimal) measure by which to determine the maximum eccentricity for which our analytic model can be trusted. This maximum eccentricity, of course, varies with the detector and source considered, but typically the eccentricity threshold is around $0.14$--$0.22$ for ground-based sources when considering comparable mass inspirals, and $10^{-3}$ for space-based detectors when considering intermediate mass-ratio inspirals.

The accuracy of the PC model deteriorates faster with initial eccentricity for space-based detectors because the theory we chose to study forces us to consider only intermediate mass-ratio inspirals, which are much more sensitive to the details of the modeling and the PN truncation of the series, as shown in~\cite{Moore:2019xkm}. In a large class of ST theories (including Jordan-Brans-Dicke-Fierz theory), the no-hair theorems have been shown to apply~\cite{Hawking:1972qk,Sotiriou:2011dz}, which then imply ST black holes are identical to those in GR. Therefore, the best tests of ST with GWs come from considering mixed systems, a BH-NS binary\footnote{The dominant modification in ST theories (dipole radiation) is suppressed in NS-NS binaries, because NSs have similar sensitivities.~\cite{Will:2010gnb,Eardley75,Will77,Will+Zag89}}.  For ground-based detectors, we can consider BH-NS binaries with somewhat comparable mass-ratios, since the total mass of the system would still be low enough for the inspiral to be in their sensitivity band. For space-based detectors, however, we must consider BH-NS binaries where the BH component is quite massive (total masses larger than $10^{2} M_{\odot}$); alternatively one can consider white dwarf-NS binaries, but these sources are barely chirping in frequency, and thus, constraints are more challenging~\cite{Littenberg:2018xxx}. These considerations, in turn, force us to consider intermediate mass-ratio inspirals, whose accuracy is much more sensitive to the details of the modeling than comparable-mass inspirals, as found e.g. in~\cite{Yunes:2008tw,Yunes:2009ef}. 

\begin{figure}[htb]
\includegraphics[width=10cm,clip=true]{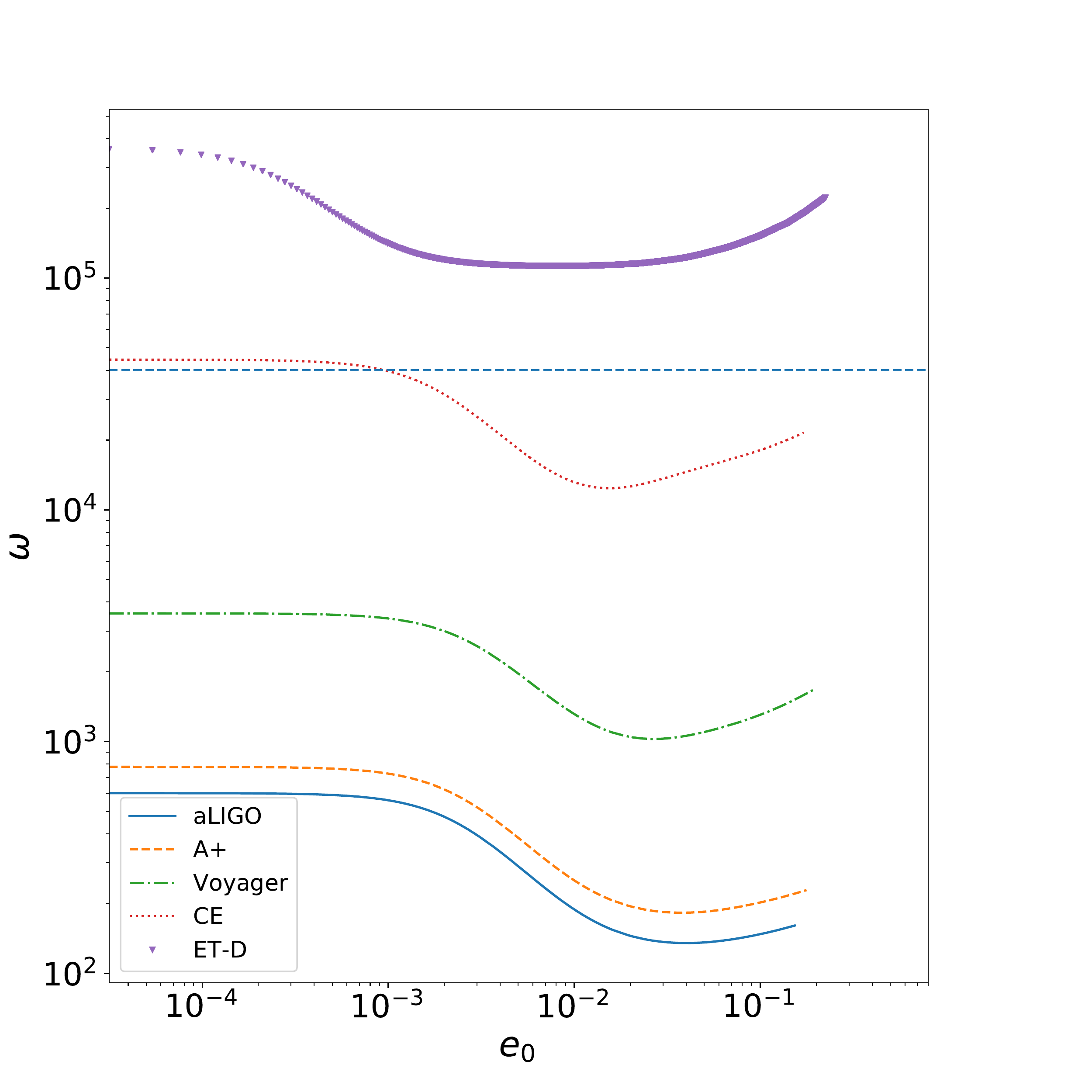}
  \caption{(Color Online) Projected constraint on $\omega$ as a function of initial eccentricity of a BH-NS signal with component masses $(10M_\odot,1.4M_\odot)$ and at a fixed luminosity distance of $D_{L} = 100$ Mpc, for a variety of ground-based detectors. The horizon dashed line is the best current constraint on $\omega$ from the tracking of the Cassini spacecraft~\cite{Bertotti03}. Observe that initially the constraint deteriorates, and eventually it improves, as the eccentricity increases, with the best constraints achievable with CE and ET.}
 \label{fig:omega-ground}
\end{figure}

With this at hand, we estimate the accuracy with which we would be able to constrain Jordan-Brans-Dicke-Fierz theory, given future observations consistent with Einstein's predictions. This estimate, shown in Fig.~\ref{fig:omega-ground} for a particular binary, is obtained through a Fisher analysis of a sky-averaged version of the analytic waveform model we develop in this paper. As expected, our Fisher results reduce to the quasi-circular ones for initial eccentricities below $10^{-4}$. In the quasi-circular limit, the constraints become more stringent with detector upgrade because we keep the luminosity distance to the source fixed (at 100 Mpc), which has the effect of increasing the signal-to-noise ratio with detector upgrade. The projected constraints with ET are more stringent that those with CE because the former has better noise performance at lower frequencies (for the configurations we studied), where negative PN corrections are important.  As we increase the initial eccentricity of the signal (between $10^{-4}$ and $10^{-2}$), we discover a partial covariance between the initial eccentricity parameter and the ST coupling parameter $\omega$, which deteriorates the measurement accuracy of both by roughly a factor of three. Eventually, as we increase the initial eccentricity of the signal further (above $10^{-2}$), the partial covariance is broken, and the accuracy to which $\omega$ can be constrained improves. The maximum initial eccentricities we can model, however, are not high enough to show how much the constraint can be improved.

The results described above have several important implications for future precision tests of GR with GWs. The first conclusion is that eccentricity can deteriorate the accuracy to which modifications to GR can be constrained, due to degeneracies that emerge between eccentric effects and modified gravity effects. This result was not presented in~\cite{Yagi10} because that analysis neglected eccentricity corrections in the ST sector of the GW model. These corrections are precisely the ones that deteriorate our ability to test GR because they enter at negative PN order relative to the leading Newtonian order term in the quasi-circular limit. A second conclusion, and corollary of the first, is that the construction of eccentric waveform models (both in GR and in modified theories) that are accurate at moderate eccentricity is urgent. Signals with initial eccentricities around $0.3$--$0.6$ could lead to more stringent constraints than the ones reported here, but the only analytic model that exists to date that is capable of representing such GWs even in GR is that of~\cite{Moore:2018kvz,Moore:2019xkm}, which has only very recently been proposed.  A third conclusion is that third-generation ground-based detectors, especially those highly sensitive at low frequencies like ET and CE, as well as spaced-based detectors like LISA (shown later in Fig.~\ref{fig:lisa}), could allow constraints on ST theories an order of magnitude more stringent that Solar System ones. 

The rest of this paper presents the details of the results reported above and it is organized as follows. 
Section~\ref{sec:BD-basics} provides a discussion of the basics of compact binary inspirals in Jordan-Brans-Dicke-Fierz theory, and it derives the evolution of the eccentricity in the frequency domain. 
Section~\ref{sec:waveform} uses the PC approximation and the SPA to construct a Newtonian-accurate analytical expression for frequency-domain gravitational waveforms produced by eccentric, inspiraling compact binaries. 
Section~\ref{sec:taylort4} introduces the 3PN eccentric TaylorT4 model. 
Section~\ref{sec:overlap} calculates the overlap between these two waveforms to find the maximum initial eccentricity that the PC approximation is valid to. In order to achieve the goal, we first derive analytical formula to maximize the inner product over the phase offset in Sec. \ref{sec:overlap-max}, and then apply the result to both ground- and space-based detectors in Secs. \ref{sec:overlap-gound} and \ref{sec:overlap-lisa}. 
Section \ref{sec:fisher-matrix} carries out a Fisher analysis to investigate the behavior of projected constraints on $\omega$ as a function of eccentricity. 
Section \ref{sec:conclusions} concludes and points to future research.

Throughout this paper we use the follow conventions unless stated otherwise. We use geometric units with $G=1=c$. We denote the masses of the binary components by $m_{1,2}$, where we choose $m_1>m_{2}$. Three-dimensional vectors are denoted with a boldface.

\section{compact binary inspirals and gravitational radiation in Jordan-Brans-Dicke-Fierz theory}
\label{sec:BD-basics}

In this section, we review some basic equations of motion and of gravitational radiation for compact binary inspirals in Jordan-Brans-Dicke-Fierz theory. For the sake of conciseness, we only provide some background and some mathematical content that will be needed in later sections. We refer the interested reader to Refs. \cite{Will+Zag89,Eardley75,Will77,Will94} for further details. All equations are shown to lowest Newtonian order in a PN expansion for simplicity, although models in later sections are extended to 3PN order.

\subsection{Conservative dynamics}
\label{sec:orbits}
Let $\bm{r}$ represent the relative vector separation between the two bodies in a binary system. The equation of motion of the binary in ST theories can then be cast as
\begin{align}
\frac{\mathrm{d}^2\bm{r}}{\mathrm{d}t^2}=-\mathcal{G}\frac{M\bm{r}}{r^3},  \label{motion-equation}
\end{align}
where $r = |\bm{r}|$, $M=m_{1}+m_{2}$ is the total mass of the binary and $\mathcal{G}$ is defined as 
\begin{align}
\mathcal{G}&=1-\xi(s_1+s_2-2s_1s_2)\,,
\end{align}
with the ST parameter
\begin{align}
\xi&=\frac{1}{2+\omega}.
\end{align}

The sensitivities $s_{A}$ represent the inertial response of the $A$th body to a change in the local value of the gravitational constant $G$. This quantity can be defined via
\begin{align}
s_{A}\equiv-\frac{\partial\ln m_{A}}{\partial\ln G}\,, \label{sensitivity-definition}
\end{align}
which in the weak-field limit reduces to the gravitational self-energy of the body, i.e.~its compactness. For neutron stars, $s$ depends on the equation of state, the relation between the internal pressure and the interior density of the compact object. In this paper, we use the APR equation of state \citep{Akmal98} as a representative example. As pointed out by Eardley \citep{Eardley75}, in the general-relativistic limit $\omega\to\infty$, Eq. (\ref{sensitivity-definition}) can be approximated by
\begin{align}
s=\frac{3}{2}\left[1-\frac{N}{m}\left(\frac{\mathrm{d}m}{\mathrm{d}N}\right)_G\right], \label{sensitivity-approximation}
\end{align}
where $N$ is total baryon number. For black holes, $s \equiv 0.5$ by the no-hair theorems, since then $m\propto G^{-1/2}$. The dependence of the inspiral motion on the sensitivities is sometimes considered to be ``smoking-gun'' evidence for a violation of the strong equivalence principle.

The equation of motion in Eq. (\ref{motion-equation}) takes the same form as that of Newtonian mechanics, with all ST corrections absorbed in $\mathcal{G}$. We can thus directly write down Kepler's third law for a binary system with orbital period $P$ and semi-major axis $a$:
\begin{align}
\frac{P}{2\pi}=\sqrt{\frac{a^3}{\mathcal{G}M}}, \label{kepler}
\end{align}
The conserved energy $E$ and angular momentum $L$ of the binary is then  
\begin{align}
E&=-\frac{\mathcal{G}M\mu}{2a}, \label{energy}\\
L&=\mu\sqrt{\mathcal{G}Ma(1-e^2)}, \label{angular}
\end{align}
where $e$ is the eccentricity of the orbit.

\subsection{Dissipative dynamics}
\label{sec:radiation}
Compact binary systems lose energy and angular momentum due to gravitational radiation leading to a quasi-circular inspiral. The rate of loss to lowest order in $1/\omega$ is \cite{Peters64,Will+Zag89,Loutrel14}.
\begin{align}
&\dot{E}=-\frac{32}{5}\frac{M^5\eta^2}{a^5}\frac{1+\frac{73}{24}e^2+\frac{37}{96}e^4}{(1-e^2)^{7/2}}-\frac{2}{3}\frac{\mathcal{S}^{2}M^4\eta^2}{a^4\omega}\frac{1+\frac{1}{2}e^2}{(1-e^2)^{5/2}}, \\
& \dot{L}=-\frac{32}{5}\eta^2\frac{M^{9/2}}{a^{7/2}(1-e^2)^2}\left(1+\frac{7}{8}e^2\right)-\frac{2}{3}\frac{\mathcal{S}^2\eta^2M^{7/2}}{\omega a^{5/2}(1-e^2)},
\end{align}
where $\mathcal{S}=s_1-s_2$ is the sensitivity difference. Clearly, these expression reduce to the GR limit in the $\omega \to \infty$ limit and all modifications are proportional to $1/\omega$ when $\omega \gg 1$.

Using Kepler's third law in Eq. (\ref{kepler}), together with the flux expressions in Eqs. (\ref{energy}) and (\ref{angular}), one can obtain evolution equations for the mean orbital frequency $F \equiv 1/P$  and orbital eccentricity
\begin{align}
\dot{F}&=\frac{48}{5\pi\mathcal{M}^2}(2\pi\mathcal{M}F)^{11/3}\frac{1+\frac{73}{24}e^2+\frac{37}{96}e^4}{(1-e^2)^{7/2}} \notag \\
&+\frac{48}{5\pi}b\mathcal{M}\eta^{2/5}(2\pi F)^{3}\frac{1+\frac{1}{2}e^2}{(1-e^2)^{5/2}}, \label{Fdot}\\
\dot{e}&=-\frac{1}{15}\mathcal{M}^{5/3}(2\pi F)^{8/3}\frac{304+121e^2}{(1-e^2)^{5/2}}e \notag \\
&-\frac{48}{5}b\eta^{2/5} \mathcal{M}(2\pi F)^2\frac{e}{(1-e^2)^{3/2}}, \label{edot}
\end{align}
where we define a new quantity $b$ following the conventions of \citeauthor{Will+Zag89} \citep{Will+Zag89}
\begin{align}
b\equiv\frac{5}{48}\frac{\mathcal{S}^2}{\omega}. \label{b-s}
\end{align}
Solving these two differential equations gives the evolution of $e$ and $F$ in the time domain. 

Most data analysis, however, is  performed in the frequency domain, and thus, it is important to express the eccentricity as a function of the orbital frequency, i.e., $e(F)$. This can be obtained from the chain rule, $\mathrm{d}e/\mathrm{d}F=(\mathrm{d}e/\mathrm{d}t) (\mathrm{d}F/\mathrm{d}t)^{-1}$, leading to 
\begin{align}
\frac{\mathrm{d} \zeta}{\mathrm{d}e}=&-\frac{3(96+292e^2+37e^4)}{e(1-e^2)(304+121e^2)} \zeta\notag \\
&-\frac{1440(32-33e^2+e^4)}{e(304+121e^2)^2} \tilde{b} \zeta^{1/3}+\mathcal{O}( \tilde{b}^2),
\end{align}
where we have defined two new quantities $\zeta \equiv2\pi F\mathcal{M}$ and $\tilde{b} \equiv b\eta^{2/5}$, and expanded in $b \ll 1\rightarrow \tilde{b} \ll 1$. 

This equation can be solved perturbatively. Consider the ansatz $\zeta=\zeta^{(0)}+\zeta^{(1)}\tilde{b}+\mathcal{O}(\tilde{b}^2)$, so that 
\begin{align}
\frac{\mathrm{d} \zeta^{(0)}}{\mathrm{d}e}=-\frac{3(96+292e^2+37e^4)}{e(1-e^2)(304+121e^2)} \zeta^{(0)},
\end{align}
\begin{align}
\frac{\mathrm{d} \zeta^{(1)}}{\mathrm{d}e}&=-\frac{3(96+292e^2+37e^4)}{e(1-e^2)(304+121e^2)} \zeta^{(1)} \notag\\
&-\frac{1440(32-33e^2+e^4)}{e(304+121e^2)^2}\left(\zeta^{(0)}\right)^{1/3}.
\end{align}
Solving these equations, we find
\begin{align}
\zeta^{(0)}&=c_0\left[\frac{1-e^2}{(1+\frac{121}{304}e^2)^{870/2299}e^{12/19}}\right]^{3/2}=c_0\sigma(e)^{-3/2}, \\
\zeta^{(1)}&=c_1\sigma(e)^{-3/2}+c_0^{1/3}\sigma(e)^{-3/2}G(e),
\end{align}
where $c_0$ and $c_1$ are integration constants and $G(e)$ is defined as
\begin{align}
G(e)&\equiv\frac{3e^{12/19}}{1520}\left[3e^2 \; {}_2F_1\left(\frac{25}{19},\frac{3728}{2299},\frac{44}{19},-\frac{121}{304}e^2\right)\right. \notag \\
&\left.-400 \; {}_2F_1\left(\frac{6}{19},\frac{3728}{2299},\frac{25}{19},-\frac{121}{304}e^2\right)\right],
\end{align}
with ${}_2F_1$ a hypergeometric function. At ${\cal{O}}(\tilde{b}^0)$, one clearly recovers the GR result \citep{Yunes09}.

The complete solution is obtained by determining the constants of integration from the initial conditions chosen for the evolution of the orbital frequency.  As is typical in perturbation theory, we choose 
\begin{align}
&\zeta^{(0)}(e_0)=\zeta_0, \qquad
\zeta^{(1)}(e_0)=0,
\end{align} 
where the quantity $e_0$ is defined as the eccentricity when the binary is at some orbital frequency. Henceforth, we define $e_{0}$ as the orbital eccentricity at the orbital frequency $F_{0}$. In the PC limit (i.e.~for very small eccentricities), $e_{0}$ also corresponds to the eccentricity at the frequency at which the (dominant mode of the) GW signal enters the detector sensitivity band\footnote{For moderate or high eccentricity signals, however, since $e_{0}$ is defined in terms of the orbital frequency, it cannot be identified with a single GW frequency.}. The eccentricity $e_{0}$ is related to $\zeta_0$ via $F_0= \zeta_0/(2\pi\mathcal{M})$. With this at hand, the complete solution is then
\begin{align}
\zeta=&\zeta_0\sigma(e_0)^{3/2}\sigma(e)^{-3/2} \notag \\
&+\tilde{b}\zeta_0^{1/3}\sigma(e_0)^{1/2}\sigma(e)^{-3/2}[G(e)-G(e_0)] \label{xi-e}
\end{align}
In the small eccentricity limit, $G(e)$ can be expanded as 
\begin{align}
G(e)\sim &-\frac{15}{19}e^{12/19}\left(1-\frac{23451}{144400}e^2+\frac{1116303}{22936496}e^4 \right. \notag \\
&\left.-\frac{1185957185}{72262473216}e^6+\frac{1617933701811}{278258696863744}e^8+\mathcal{O}(e^{10})\right),
\end{align}
which then provides an expression for the orbital frequency $F$ as a function of the eccentricity $e$.

The eccentricity as a function of orbital frequency is obtained by inverting Eq. (\ref{xi-e}), which can be decomposed into a GR term and a Jordan-Brans-Dicke-Fierz term:
\begin{align}
e(F) = e_{\GR}(F)+e_{\BD}(F), \label{e-f}
\end{align}
where $e_{\GR}(F)$ is given in~\cite{Yunes09} to ${\cal{O}}(e_{0}^{8})$, while
\begin{widetext}
\begin{align}
e_{\BD}(F)&=\tilde{b}e_0\chi^{-19/18}\zeta_0^{-2/3}\left[\frac{5}{6}(1-\chi^{-2/3})+e_0^2\left(\frac{23201}{54720}-\frac{16615}{10944}\chi^{-2/3}+\frac{103033}{18240}\chi^{-25/9}-\frac{16615}{3648}\chi^{-19/9}\right)\right. \notag \\
&+\left.e_0^4\left(\frac{803477993}{2195804160}-\frac{4398413525}{2195804160}\chi^{-2/3}+\frac{342378659}{11089920}\chi^{-25/9}-\frac{69912597}{3696640}\chi^{-19/9}-\frac{91775102957}{2195804160}\chi^{-44/9}\right.\right. \notag \\
&\left.\left.+\frac{1256493575}{39923712}\chi^{-38/9}\right)+e_0^6\left(\frac{3691296108661}{12015440363520}-\frac{528671699005}{218462552064}\chi^{-2/3}+\frac{1307794056779}{13485342720}\chi^{-25/9} \right.\right.\notag \\ 
&\left.\left.-\frac{469302381929}{9889251328}\chi^{-19/9}-\frac{304968667126111}{801029357568}\chi^{-44/9}+\frac{17867388896243}{72820850688}\chi^{-20/9}+\frac{778111645240871}{2403088072704}\chi^{-7} \right.\right. \notag \\
&\left.\left.-\frac{51523706370955}{218462552064}\chi^{-19/3}\right)\right],
\end{align}
\end{widetext}
with $\chi=\zeta/\zeta_0=F/F_0$. This equation guarantees that $e_{\BD}(F=F_{0}) = 0$, and thus, $e(F_0)=e_0$. Clearly then, the ST modification $e_{\BD}$ depends on $e_0$, $F_0$ and $\chi$, while the GR term $e_{\GR}$ only depends on $e_0$ and $\chi$. 

\section{Gravitational Wave Models For Eccentric Inspirals in Jordan-Brans-Dicke-Fierz Theory}
\label{sec:waveform}
In this section, we discuss how to construct GW models that will be used in later sections. Two types of waveforms are constructed. In Sec. \ref{section:spa}, we construct an analytical, frequency-domain, GW model for eccentric inspirals in Jordan-Brans-Dicke-Fierz theory within the PC approximation introduced in~\citep{Yunes09}. In Sec. \ref{sec:taylort4}, we describe how to construct a 3PN accurate eccentric time-domain model, an analog to the TaylorT4 model but for eccentric binaries in GR, and discuss details of its discrete Fourier transform (DFT).

\subsection{Analytic model}
\label{section:spa}
We begin with a brief review of the PC approximation to compute analytic, frequency-domain waveforms for eccentric inspirals. The plus and cross polarizations, $h_+$ and $h_\times$, can be written as a sum over harmonics of the orbital phase $\phi$. In eccentric binaries, this quantity is not simply the product of the angular velocity and time, but rather, it is related to the mean anomaly $l = n \, t = (2 \pi/P) t$, where $n$ is the mean motion and $P$ is the orbital period, via
\begin{align}
\cos\phi=&=-e+\frac{2}{e}(1-e^2)\sum_{\ell=1}^{\infty}J_\ell(\ell e)\cos \ell l, \label{cosphi-l} \\
\sin\phi=&(1-e^2)^{1/2}\sum_{k=1}^{\infty}[J_{\ell-1}(\ell e)-J_{\ell+1}(\ell e)]\sin \ell l, \label{sinphi-l}
\end{align} 
where $J_\ell(\ell e)$ is the Bessel function of the first kind and $e$ is the orbital eccentricity.

Using Eqs. (\ref{cosphi-l}) and (\ref{sinphi-l}) in the harmonic decomposition of the waveform polarizations, one can write
\begin{align}
h_{+,\times}=\mathcal{A}\sum_{\ell=1}^{10}[C_{+,\times}^{(\ell)}\cos\ell l+S_{+,\times}^{(\ell)}\sin\ell l]\,,
\end{align}
where $C_{+,\times}^{(\ell)}$ and $S_{+,\times}^{(\ell)}$ are polynomials of $e$, whose coefficients are trigonometric functions of the inclination and the polarization angles $\iota$ and $\beta$~\citep{Martel99} (see Appendix B of \cite{Yunes09}). We have here truncated the sums at $\ell = 10$ so as to obtain expressions accurate to $\mathcal{O}(e^8)$~\cite{Yunes09}.  Technically, the ST polarizations will have additional contributions from PN corrections to the amplitude of the expression provided above\footnote{The metric perturbation also contains a propagating scalar (breathing) mode, but this term will not be included here because it is not directly detectable with only 2 interferometers~\cite{Chatziioannou:2012rf}.}, but we neglect those here because we are searching for a waveform model to leading order in the GR deformation.

With this at hand, we can now compose the response function, the time-domain strain measured by detectors in response to an impinging GW, to find
\begin{align}
h(t)&=F_+(\theta_S,\phi_S,\psi_S)h_++F_\times(\theta_S,\phi_S,\psi_S)h_\times \notag \\
&=\mathcal{A}\sum_{\ell=1}^{10}\alpha_\ell\cos(\ell l+\phi_\ell), \label{pattern}
\end{align}
where $\alpha_\ell=\text{sgn}(\Gamma_\ell)\sqrt{\Gamma_\ell^2+\Sigma_\ell^2}$ and $\phi_\ell=\arctan\left(-\frac{\Sigma_\ell}{\Gamma_\ell}\right)$ are functions of $\Gamma_\ell=F_+C_+^{\ell}+F_\times C_\times^{\ell}$ and $\Sigma_\ell=F_+S_+^{\ell}+F_\times S_\times^{\ell}$. The beam pattern functions $F_{+,\times}(\theta_S,\phi_S,\psi_S)$ depend on the location of source relative to the detector through the sky angles $\theta_S$ and $\phi_S$, as well as on a polarization angle $\psi_S$.

The Fourier transform of the time-domain response in Eq.~\eqref{pattern} can be modeled in the SPA. In the latter, one expands the Fourier integral in the ratio of the radiation-reaction time scale to the orbital period, keeping terms of leading order in this ratio; higher-order terms are subdominant and can be neglected~\citep{Droz99}. The SPA to the Fourier transform of the response function is
\begin{align}
\tilde{h}(f)&=-\left(\frac{5}{384}\right)^{1/2}\pi^{-2/3}\frac{\mathcal{M}^{5/6}}{D_{L}}f^{-7/6}\notag \\
&\times\sum_{\ell=1}^{10} \varpi_{\ell}\left(\frac{\ell}{2}\right)^{2/3}e^{-i(\pi/4+\Psi_{\ell})}{ \Theta(\ell F_\text{ISCO}-f)}, \label{spa-waveform}
\end{align}
where $f$ is the Fourier frequency, $D_{L}$ is the luminosity distance, and $\mathcal{M}\equiv M\eta^{3/5}$ is the chirp mass, with $\eta\equiv \mu/M = m_1m_2/M^2$ the symmetric mass ratio and $\mu$ the reduced mass. We truncate the waveforms with unit step functions $\Theta(x)$ to make sure each harmonic does not exceed its region of validity (see e.g. the discussion in Appendix~A of~\cite{Moore:2018kvz}).

The quantity $\varpi_\ell$ arises due to the $\dot{F}^{-1/2}$ factor that is in the amplitude of the SPA. In Jordan-Brans-Dicke-Fierz theory, the rate of change of the orbital frequency is governed by Eq.\eqref{Fdot}, and thus, one finds
\begin{align}
\varpi_{\ell}&\!=\!\!\left[\frac{1+\frac{73}{24}e^2+\frac{37}{96}e^4}{(1-e^2)^{7/2}}\!+ \tilde{b} \left(\frac{2\pi f\mathcal{M}}{\ell}\right)^{-2/3} \!\!\! \frac{1+\frac{1}{2}e^2}{(1-e^2)^{5/2}}\right]^{-\frac{1}{2}}
\nn \\
&\times \alpha_{\ell}e^{-i\phi_{\ell}}. \label{xi-BD}
\end{align}
This expression reduces to that of \citep{Yunes09} in the GR limit, when $\tilde{b} \to 0$. The coefficients $\varpi_\ell$ should be re-expanded in $e_0\ll 1$ to $\mathcal{O}(e_0^8)$, using the expressions of $\Gamma_\ell$ and $\Sigma_\ell$, as well as $e(F)$ in Eq. (\ref{e-f}); in the GR limit, such re-expansion was presented in Appendix C of Ref. \citep{Yunes09}, with $\beta$ and $\iota$ fixed set to zero. We provide similar expressions for $\varpi_\ell$ in ST theories as functions of $e(F)$ (but without re-expanding in $e_{0} \ll 1$ in Appendix \ref{sec:app-xi}, with $\iota=\beta=0$.

Let us now turn to the calculation of the Fourier phase $\Psi_\ell$. In the SPA, this phase can be expressed as
\begin{align}
\Psi_\ell=-2\pi f t_s+ \ell \; l(t_s), \label{gw-phase-spa}
\end{align}
where $t_s$ is the stationary point and $l(t_{s})$ is the time-domain mean anomaly evaluated at $t_{s}$. The stationary point is defined via the stationary phase condition $F(t_s)=f(t_s)/\ell$, where $F(t)$ is the mean orbital frequency, i.e.~the time derivative of the mean anomaly. The orbital phase and the stationary point can be written as 
\begin{align}
&l(t_s)=l_c+2\pi\int_\text{coalescence}^{F(t_s)}\tau^\prime\mathrm{d}F^\prime, \label{phase-spa} \\
&t(t_s)=t_c+\int_\text{coalescence}^{F(t_s)}\frac{\tau^\prime}{F^\prime}\mathrm{d}F^\prime, \label{time-spa}
\end{align}
where we have defined  $\tau\equiv F/\dot{F}$, and where $l_c$ and $t_c$ are the mean anomaly at coalescence and the time of coalescence, respectively, i.e.~the orbital phase and time at which the orbital frequency diverges. 

In order to evaluate these two integrals, we need to first evaluate $\tau$ using Eqs. (\ref{e-f}) and Eq. (\ref{Fdot}) and then expand this quantity in $e_{0} \ll 1$ to $\mathcal{O}(e_{0}^8)$ and in $b \ll 1$. Keeping terms up to $\mathcal{O}(e_0^8)$ and $\mathcal{O}(b^1)$, we find
\begin{align}
\tau=\tau_{\GR}+\tau_{\BD},
\end{align}
where $\tau_{\GR}$ is given in Eq. (4.27) of~\cite{Yunes09} and $\tau_{\BD}$ is 
\begin{widetext}
\begin{align}
\tau_{\BD}=&-\frac{5}{96} \tilde{b} \; \mathcal{M} \; (2\pi\mathcal{M}F)^{-10/3}\left[1+\left(\frac{785}{72}\chi^{-13/9}-\frac{1511}{72}\chi^{-19/9}\right)e_0^2+\left(\frac{87685679}{328320}\chi^{-38/9}-\frac{5222765}{32832}\chi^{-32/9}\right.\right.\notag \\
&\left.\left.
-\frac{5021053}{65664}\chi^{-19/9}+\frac{4171333}{164160}\chi^{-13/9}\right)e_0^4+\left(-\frac{30003281383}{10165760}\chi^{-19/3}+\frac{678205125}{369664}\chi^{-17/3}+\frac{291379511317}{149713920}\chi^{-38/9}\right.\right.\notag \\
&\left.\left.-\frac{142281697789}{149713920}\chi^{-32/9}-\frac{2553265135}{14971392}\chi^{-19/9}+\frac{67686858773}{1646853120}\chi^{-13/9}\right)e_0^6+\left(\frac{55119048817407295}{1802316054528}\chi^{-76/9}\right.\right. \notag \\
&\left.\left.-\frac{794117334022775}{40961728512}\chi^{-70/9}-\frac{99700904035709}{3090391040}\chi^{-19/3}+\frac{12614342351649529}{873850208256}\chi^{-17/3}+\frac{2153613676818511}{273078190080}\chi^{-38/9}\right.\right. \notag \\
&\left.\left.+\frac{2805180056710151}{873850208256}\chi^{-11/3}-\frac{76111514161609}{23467656960}\chi^{-32/9}-\frac{12521507252345}{40961728512}\chi^{-19/9}+\frac{260148130266001}{4505790136320}\chi^{-13/9}\right)e_0^8\right] \label{tau-BD}
\end{align}
where recall that $\chi := F/F_{0}$ and $e_{0}$ is the eccentricity at frequency $F_{0}$.

With this at hand, we can now compose the Fourier phase in the SPA. Defining the quantity $x=(\pi\mathcal{M}f)^{5/3}$ for consistency with \cite{Yunes09}, and combining Eqs. (\ref{gw-phase-spa}), (\ref{phase-spa}), (\ref{time-spa}) and (\ref{tau-BD}), we find
\begin{align}
&\Psi_{\ell}=-2\pi ft_{c}
+\ell l_c+\frac{3}{128x}\left(\frac{\ell}{2}\right)^{8/3}\Xi_{\ell} \label{tc-phic}\,,
\end{align}
where $\Xi_{\ell} := \Xi^\text{PC}_{\ell}+\Xi^{\BD}_{\ell}$, with $\Xi^\text{PC}_{\ell}$ given in Eq. (4.28) of Ref. \cite{Yunes09} and $\Xi^{\BD}_{\ell}$ given by
\begin{align}
\Xi^{\BD}_{\ell}&=\frac{1}{2} \; \tilde{b} \; (2\pi f_0\mathcal{M}/\ell)^{-2/3}\left[\frac{8}{7}\chi_{\ell}^{-2/3}+\left(\frac{3925}{731}\chi_{\ell}^{-19/9}-\frac{1511}{196}\chi_{\ell}^{-25/9}\right)e_0^2+\left(\frac{87685679}{1829472}\chi_{\ell} ^{-44/9}-\frac{26113825}{749208 }\chi_{\ell} ^{-38/9} \right.\right. \notag \\
&\left.\left.-\frac{5021053}{178752}\chi_{\ell}^{-25/9}+\frac{4171333}{333336} \chi_{\ell}^{-19/9}\right)e_0^4+\left(-\frac{30003281383}{95812288}\chi_{\ell}^{-7}+\frac{376780625}{1663488}\chi_{\ell}^{-19/3}+\frac{291379511317}{834239232}\chi_{\ell}^{-44/9} \right. \right. \notag \\
&\left.\left.-\frac{142281697789}{683277696}\chi_{\ell}^{-38/9}-\frac{2553265135}{40755456}\chi_{\ell}^{-25/9}+\frac{67686858773}{3344026752}\chi_{\ell}^{-19/9}\right)e_0^6+\left(\frac{275595244087036475}{128690372726784}\chi_{\ell}^{-82/9}\right.\right.\notag \\
   &\left.\left.-\frac{22689066686365}{14791735296} \chi_{\ell}^{-76/9}-\frac{99700904035709}{29126935552}\chi_{\ell}^{-7}+\frac{63071711758247645}{35390933434368}\chi_{\ell}^{-19/3}+\frac{2153613676818511}{1521652359168} \chi_{\ell}^{-44/9} \right. \right. \notag \\
& \left.\left. +\frac{14025900283550755}{20644711170048}\chi_{\ell}^{-13/3}-\frac{76111514161609}{107103778848} \chi_{\ell}^{-38/9}-\frac{1788786750335}{15929561088}\chi_{\ell}^{-25/9}+\frac{260148130266001}{9149257193472} \chi_{\ell}^{-19/9}\right)e_0^8\right]. \label{phase}
\end{align} 
\end{widetext}
with $\chi_{\ell} = f/(\ell F_{0})$
after applying the stationary phase condition, which ensures\footnote{Note that in Eq. (4.28) of Ref. \cite{Yunes09} the $\chi$ that appears in that equation should really be $\chi_{\ell}$ as defined in this paper.} 
$e(F_{0}) = e_{0}$. Observe that the ST modification to the Fourier phase contains terms that scale as $\chi_{\ell}^{-2/3}$ relative to the GR contributions; these are -1PN corrections to the GR phase, as expected from the presence of dipole radiation in the binary. Observe also that the ST modification is always proportional to $\tilde{b} = b \eta^{2/5}$, which means that $b$ and $\eta$ are completely degenerate at Newtonian order; fortunately, the symmetric mass ratio appears also at 1 PN order in the GR sector of the Fourier phase, and thus, it can be measured independently from $b$ allowing us to break this degeneracy~\citep{Will94}. 

Given the degeneracy described above, a Newtonian accurate waveform model, as presented above, is not sufficient to test ST theories. We will here work in the \emph{restricted} PN approximation, in which we include higher PN order terms to the Fourier phase, keeping the amplitude at Newtonian order. Moreover, we will only add higher PN order terms to the GR sector of the Fourier phase, since the ST sector has not been fully worked out beyond Newtonian order. We thus henceforth model the Fourier phase via
\begin{align}
\Psi_\ell=&-2\pi ft_c+\ell l_c+ \Psi^{\GR}_{\ell} + e_{0}^{2} \Psi^\GRtwo_{\ell} + e_{0}^{4} \Psi^\GRfour_{\ell}\notag \\
&+ e_{0}^{6} \Psi^\GRsix_{\ell} + e_{0}^{8} \Psi^\GReight_{\ell} + \Psi^{\BD}_{\ell}\,. \label{phase-pn}
\end{align}
The term $\Psi^{\GR}_{\ell}$ is the quasi-circular expression in GR up to 3PN order, and thus, it is independent of $e_{0}$ and contains terms up to ${\cal{O}}(1/c^{6})$ that can be found in \citep{Buonanno09}. 
The term $\Psi_{\ell}^\GRtwo$ is the $e_0^2$ correction to the quasi-circular term in GR, which is known to 3PN order and thus contains terms up to ${\cal{O}}(1/c^{6})$ that can be found in \citep{Moore16}.
The terms $\Psi^\GRfour_\ell$ and $\Psi^\GRsix_{\ell}$ are the $e_0^4$ and $e_0^6$ corrections to the quasi-circular expression in GR, which are both known to 2PN order and thus contain terms up to ${\cal{O}}(1/c^{4})$ that can be found in \citep{Tanay16}. 
The term $\Psi^\GReight_{\ell}$ is the $e_0^8$ correction to the quasi-circular term in GR, which is known only to Newtonian order and can be found in \citep{Yunes09}. 
The explicit expressions for each of these terms are also presented in Appendix \ref{sec:phase-GR-PN}.
Finally, the term $\Psi^{\BD}_{\ell}={3}/({128x}) ({\ell}/{2})^{8/3}\Xi^{\BD}_{\ell}$ contains all of the ST modifications to the GR Fourier phase up to ${\cal{O}}(e_{0}^{8})$ in the PC expansion and to leading Newtonian order in the PN approximation. This is the most accurate (in the PN and PC sense) Fourier phase in Jordan-Brans-Dicke-Fierz theory we can construct as of the writing of this paper.


\subsection{Numerical model}
\label{sec:taylort4}

In the next section, we estimate roughly the maximum initial eccentricity the previous analytical model is valid to. This will be achieved by comparing the analytic model in the GR limit to a numerical eccentric model in GR. In this subsection, we will detail the construction of the latter to 3PN order. 

Let us begin with a brief discussion of PN expansion parameters. Generally speaking, quantities related to elliptical orbit are most naturally expanded in terms of the radial orbit angular frequency $\omega_r\equiv n\equiv\xi/M$, which is nothing but the mean motion. For quasi-circular orbits, on the other hand, quantities are most naturally expanded in terms of the azimuthal or $\phi$-angular frequency $\omega_\phi\equiv\xi_\phi/M$. In this paper, we will use $\xi_\phi$ as our expansion parameter because quantities expressed in terms of this parameter have a simpler functional dependence on the orbital frequency $F$, where recall that $\xi_\phi=2\pi M F$. These two expansion parameters are related via \citep{Moore16}.
\begin{align}
\xi&=\xi_\phi\left(1-\frac{3}{1-e_t^2}\xi_\phi^{2/3}-[18-28\eta+(51-26\eta)e_t^2]\right. \notag \\
& \times\frac{\xi_\phi^{4/3}}{4(1-e_t^2)^2}-\left\{-192-(14624-492\pi^2)\eta+896\eta^2\right.\notag \\
&+[8544-(17856-123\pi^2)\eta+5120\eta^2]e_t^2+(2496-1760\eta \notag \\
&\left.\left.+1040\eta^2)e_t^4+[1920-768\eta+(3840-1536\eta)e_t^2]\right.\right. \notag \\
&\left.\left.\times\sqrt{1-e_t^2}\right\} \frac{\xi_\phi^2}{128(1-e_t^2)^3}\right)\,.
\label{xi-xiphi}
\end{align}
The expression above depends on the so-called ``temporal'' eccentricity $e_{t}$, which differs from the radial or azimuthal eccentricities starting at 2PN order. In the previous section we presented expressions at Newtonian order for the most part, so it was not necessary to differentiate among these different eccentricity parameters. When constructing an analytic model to 3PN order in the previous section, however, we do use the temporal eccentricity as our measure of eccentricity in the binary.

The eccentric TaylorT4 model we develop here requires the temporal evolution of the orbital frequency $F$ and the temporal eccentricity $e_t$. To 3PN order, the evolution equations contain instantaneous and hereditary contributions, namely
\begin{align}
\left.\frac{\mathrm{d}F}{\mathrm{d}t}\right|_\text{inst}&=\frac{\eta\xi_\phi^{11/3}}{2\pi M^2}[\mathcal{O}_\text{N}+\xi_\phi^{2/3}\mathcal{O}_\text{1PN}+\xi_\phi^{4/3}\mathcal{O}_\text{2PN}+\xi_\phi^{2}\mathcal{O}_\text{3PN}],\label{fdot-taylort4-inst}\\
\left.\frac{\mathrm{d}e_t}{\mathrm{d}t}\right|_\text{inst}&=-\frac{e_t\eta\xi_\phi^{8/3}}{M}[\mathcal{E}_\text{N}+\xi_\phi^{2/3}\mathcal{E}_\text{1PN}+\xi_\phi^{4/3}\mathcal{E}_\text{2PN}+\xi_\phi^{2}\mathcal{E}_\text{3PN}],\label{edot-taylort4-inst}
\end{align}
and 
\begin{align}
\left.\frac{\mathrm{d}F}{\mathrm{d}t}\right|_\text{hered}&=\frac{48}{5\pi}\frac{\eta}{M^2}\xi_\phi^{11/3}[\xi_\phi\mathcal{A}_\text{1.5PN}+\xi_\phi^{5/3}\mathcal{A}_\text{2.5PN}+\xi_\phi^{2}\mathcal{A}_\text{3PN}],\label{fdot-taylort4-hered}\\
\left.\frac{\mathrm{d}e_t}{\mathrm{d}t}\right|_\text{hered}&=\frac{32}{5}\frac{e_t\eta}{M}\xi_\phi^{8/3}[\xi_\phi\mathcal{K}_\text{1.5PN}+\xi_\phi^{5/3}\mathcal{K}_\text{2.5PN}+\xi_\phi^{2}\mathcal{K}_\text{3PN}],\label{edot-taylort4-hered}
\end{align}
which can be found in Eqs. (6.14), (6.18), (6.24c) and (6.25) of Ref.~\citep{Arun09}. In that paper, however, these equations are expressed in ADM coordinates, while in our paper we use modified harmonic coordinates. The coordinate-transformed expressions can be obtained by substituting Eq. (4.15) of Ref. \citep{Arun09} into Eqs. (\ref{fdot-taylort4-inst}), (\ref{edot-taylort4-inst}), (\ref{fdot-taylort4-hered}) and (\ref{edot-taylort4-hered}). We should note that only the instantaneous parts need to be transformed. The hereditary contributions remain the same up to the 3PN order in both coordinates. Although the explicit form of the angular-momentum flux in MH coordinates is shown in Appendix C of Ref. \citep{Arun09}, to our knowledge the explicit form of $\dot{F}$ and $\dot{e}$ had not previously appeared in the literature before, so we present them in Appendix \ref{sec:Fdot-edot-MH}.

We obtain the temporal evolution of $F$ and $e_{t}$ by numerically solving the two differential equations presented above. We choose the initial conditions 
\begin{align}
&F(t=0)= F_{0}, \label{initial}\\
& e_t(t=0)=e_0, \label{initial-e0}
\end{align}
where $F_0$ is the initial orbital frequency and $e_0$ is the corresponding initial eccentricity, as discussed in Sec. \ref{section:spa}. Note that for the very small eccentricity systems that we consider, $F_{0} \approx f_{0}/2$, where $f_{0}$ is the initial GW frequency of the $\ell=2$ harmonic, which is the dominant harmonic in the signal. We stop all numerical evolutions at the innermost stable circular orbit (ISCO) of a test particle around a Schwarzschild BH, i.e., $F(t_\text{end})=F_\text{ISCO}=\frac{1}{2\pi6^{3/2}M}$. We uniformly sample the waveforms from $t=0$ to $t=t_\text{end}$ with $N$ points and a temporal discretization $\Delta t =t_\text{end}/(N-1)$.

Once we have $F(t)$ and $e_t(t)$, we can find the mean anomaly $l$, the eccentric anomaly $u$ and the true anomaly $v$, all of which are needed to evaluate the waveform. The mean anomaly can be found by solving the differential equation
\begin{align}
\frac{\mathrm{d} l}{\mathrm{d}t}=n=\frac{2\pi F}{1+k},
\end{align}
where \citep{Moore16}
\begin{align}
k&=\frac{3\xi_\phi^{2/3}}{1-e_t^2}+[54-28\eta+(51-26\eta)e_t^2]\frac{\xi_\phi^{4/3}}{4(1-e_t^2)^2} \notag \\
&+\left\{6720-(20000-492\pi^2)\eta+896\eta^2+[18336\right. \notag \\
&-(22848-123\pi^2)\eta+5120\eta^2]e_t^2+(2496-1760\eta\notag \\
&+1040\eta^2)e_t^4+[1920-768\eta+(3840-1536\eta)e_t^2]\notag \\
&\left.\times\sqrt{1-e_t^2}\right\}\frac{\xi_\phi^2}{128(1-e_t^2)^3},
\end{align}
with the initial condition
\begin{align}
l(t=0)=0. \label{initial-l}
\end{align}
The mean anomaly is related to the eccentric anomaly $u$ by the Kepler equation, whose 3PN accurate version is given in Eq. (27) of Ref. \citep{Gopa06} in terms of $\xi$. Substituting Eq. (\ref{xi-xiphi}) into this equation, and then numerically inverting it determines $u=u(l,\xi_\phi,e_t)$. The true anomaly $v$ can be obtained from
\begin{align}
v-u=2\tan^{-1}\left(\frac{\beta_\phi\sin u}{1-\beta_\phi\cos u}\right),
\end{align}
where $\beta_\phi=(1-\sqrt{1-e_\phi^2})/e_\phi$; an explicit expression for the azimuthal eccentricity $e_\phi$ in terms of $e_t$ can be found in Eq. (3.6) of Ref. \citep{Moore16}. 

Before we can construct the waveform. we need to find the temporal evolution of the orbital phase $\phi$. This quantity can be decomposed to 3PN order via
\begin{align}
\phi=\lambda+W,
\end{align}
where $W$ is a $2\pi$-periodic function, whose 3PN analytical expressions in terms of $\xi$, $v$ and $u$ are given in Eqs. (25e)-(25h) of \citep{Gopa06}. The quantity $\lambda$ is a $2\pi(1+k)$-periodic function of the mean anomaly, which can be obtained by numerical solving the differential equation 
\begin{align}
\frac{\mathrm{d}\lambda}{\mathrm{d}t}=\omega_\phi=2\pi F,
\end{align}
with the same initial condition as the mean anomaly
\begin{align}
\lambda(t=0)=0. \label{initial-lambda}
\end{align}
Note that before solving this differential equation, one must use Eq. (\ref{xi-xiphi}) to switch PN expansion parameters.

With all of this at hand, we can now compute the waveform polarizations. Since we work in the restricted PN approximation, we here only use the leading Newtonian order expressions for the two GW polarizations \citep{Yunes09}
\begin{align}
h_+=&\frac{\mathcal{A}}{1-e_t^2}\left\{\cos\phi\left[e_ts_i^2+\frac{5}{2}e_tc_{2\beta}(1+c_i^2)\right]\right. \notag\\
&+\sin\phi\left[\frac{5e_t}{2}s_{2\beta}(1+c_i^2)\right]+\cos2\phi[2c_{2\beta}(1+c_i^2)] \notag \\
&+\sin2\phi[2s_{2\beta}(1+c_i^2)]+\cos3\phi\left[\frac{e_t}{2}c_{2\beta}(1+c_i^2)\right] \notag \\
&\left.+\sin3\phi\left[\frac{e_t}{2}s_{2\beta}(1+c_i^2)+e_t^2s_i^2+e_t^2(1+c_i^2)c_{2\beta}\right] \right\},
\end{align}
\begin{align}
h_\times&=\frac{\mathcal{A}}{1-e_t^2}\{\cos\phi[-5e_ts_{2\beta}c_i]+\sin\phi[5e_tc_{2\beta}c_i] \notag \\
&+\cos2\phi[-4s_{2\beta}c_i]+\sin2\phi[4c_{2\beta}c_i] \notag \\
&+\cos3\phi[-e_ts_{2\beta}c_i]+\sin3\phi[e_tc_{2\beta}c_i]-2e_t^2s_{2\beta}\},
\end{align}
where $c_i=\cos\iota,s_i=\sin\iota,c_{2\beta}=\cos2\beta$ and $s_{2\beta}=\sin2\beta$. The overall amplitude parameter is defined as 
\begin{align}
\mathcal{A}=-\frac{\mathcal{M}}{D_L}(2\pi\mathcal{M}F)^{2/3}.
\end{align} 
Inserting the time-dependent $F(t)$, $e_t(t)$ and $\phi(t)$ in the expressions above one can find the time-domain response function via $h(t)=F_+h_+(t)+F_\times h_\times(t)$.

Since data analysis studies are typically carried out in the frequency-domain, we need to calculate the Fourier transform of the response function. Following the procedure of Ref. \citep{Droz99}, we use a discrete Fourier transform (DFT) to do so. In particular, we will use a Fast Fourier Transform (FFT) algorithm, which requires the signal to be periodic, with period $T$, and so we first zero-pad the time-domain response on both sides:
\[
    h_\text{padding}(t)= 
\begin{cases}
    0,& 0<t<T\\
    h(t),& T<t<2T \\
    0, & 2T<t<4T\,,
\end{cases}
\]
such that the total length of the time-domain sample is $4N$. Next, we FFT the zero-padded response, $\tilde{h}(f)=\mathcal{F}[h_\text{padding}(t)]$, which returns a Fourier transform that starts at zero frequency $f=0$, with frequency interval $\Delta f=1/(4N\Delta t)$. The sample number $N$ is chosen to be large enough so that the Nyquist frequency, $f_\text{Ny}=1/(2\Delta t)$, is larger than $f_\text{ISCO}$.

\section{Validity of PC approximation}
\label{sec:overlap}

In this section, we estimate the maximum initial eccentricity that our analytic model is valid to in the GR limit. To do so, we calculate the overlap between the eccentric TaylorT4 model $\tilde{g}(f)$ and the analytic one $\tilde{h}(f)$ described in the previous section, maximized over the constant phase and time offsets $ l_c$ and $t_{c}$, as a function of initial eccentricity $e_0$. Since the analytic model is only valid in the limit of small eccentricity, the two models will dephase and the match will decrease as the initial eccentricity becomes large. In this paper, when the match drops below $97\%$ of the quasi-circular overlap, we declare the PC model invalid. Other choices to declare a model invalid are possible, and their study is relegated to future work.

\begin{figure}
\centering
\includegraphics[width=\columnwidth,clip=true]{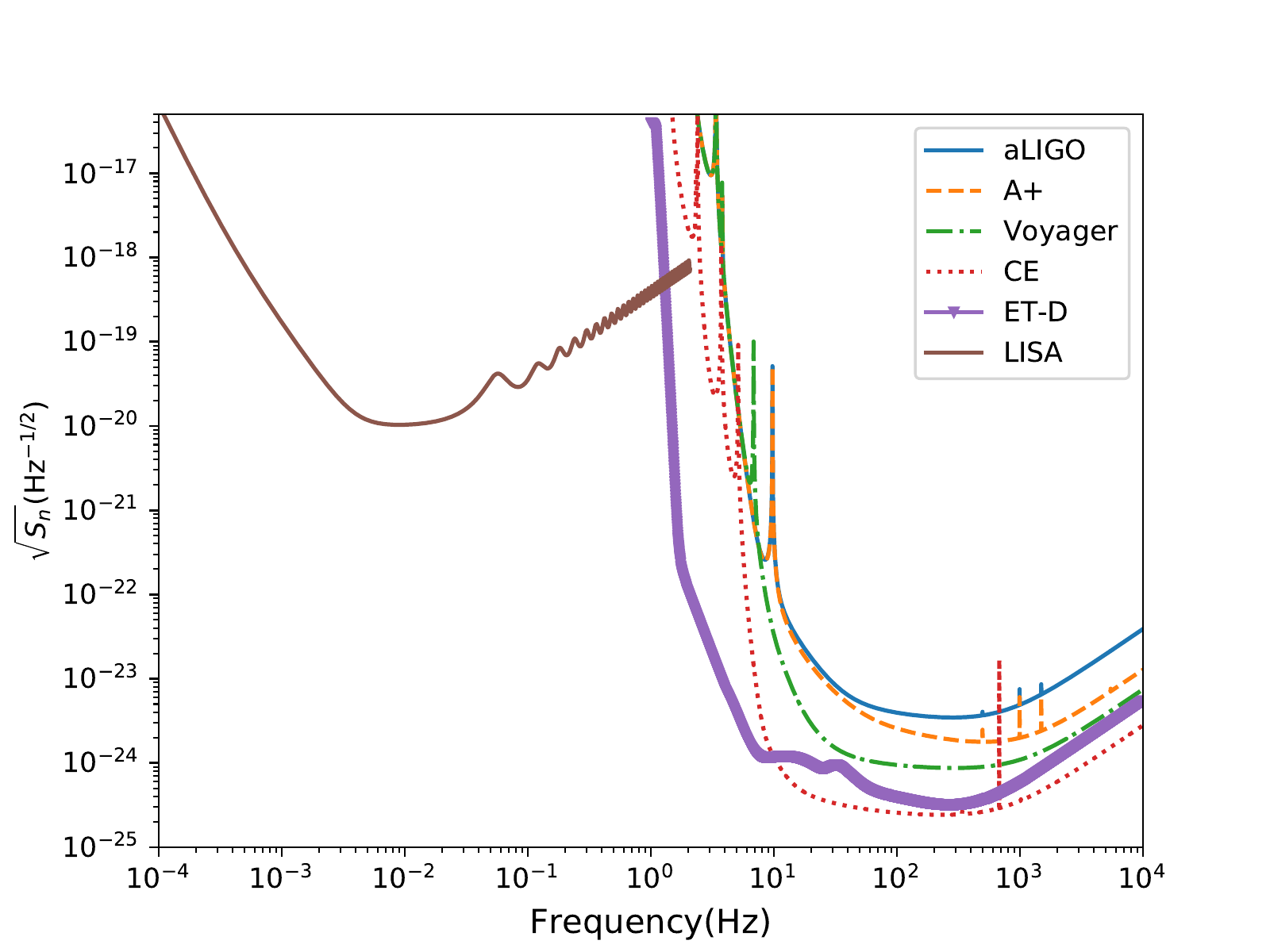}
\caption{(Color Online) Spectral noise densities of current and future ground and space-based detectors.}
\label{fig:PSD}
\end{figure}

The overlap between two waveforms $h$ and $g$ is defined in terms of their inner product, namely
\begin{align}
(h|g)=4 \; \Re \int\frac{\tilde{h}(f)^*\tilde{g}(f)}{S_n(f)}\mathrm{d}f, \label{innerproduct}
\end{align}
where the $*$ superscript is the complex conjugate operator and $S_n(f)$ is the noise spectral density of the detector. We will here consider a variety of current and future detectors, whose spectral noise density is shown in Fig. \ref{fig:PSD}~\citep{Chamberlain17,LIGOwhite,Amaro-Seoane17}. For the most part, we consider here single-detector sources, and leave a discussion of multi-band sources for future work.

The match is defined as the normalized overlap maximized over time and phase offset $t_{c}$ and $ l_c$:
\begin{align}
\mathcal{O}[h,g]=\max_{t_c,l_c}\frac{(h|g)}{\sqrt{(h|h)(g|g)}}.
\end{align}
These are ``extrinsic'' parameters that enter the analytic model as integration constants. In the quasi-circular limit, the maximization over $t_c$ can be performed through a Fourier transform trick, while the maximization of $ l_c$ can be done with two orthogonal templates \citep{Buonanno09}. This method, however, formally fails here because of the presence of multiple harmonics in the models. In this section, we will first derive analytic expressions to rapidly maximize over these extrinsic parameters in the PC approximation (see also~\cite{Moore:2018kvz}), and conclude with a study of the regime of validity in initial eccentricity. 

\subsection{Maximization over $(t_{c}, l_c)$ \\ for small eccentricity models}
\label{sec:overlap-max}

Let us begin by reviewing how this maximization is done when the model contains only a single harmonic. For a single-harmonic waveform $\tilde{h}(f)$, its SNR $\sqrt{(h|h)}$ does not depend on $t_c$ and $l_c$. Hence, maximizing the fraction in the overlap definition reduces to maximizing the inner product $(h|g)$. Suppose further that $\tilde{h}(f)$ can be written as $\tilde{h}(f)= e^{-2\pi ift_c}\tilde{h}_1(f)$; then, we have 
\begin{align}
(h|g)&=4\Re \int\frac{\tilde{h}_1(f)^*\tilde{g}(f)}{S_n(f)}e^{2\pi ift_c}\mathrm{d}f 
=4\Re\mathcal{F}^{-1}\left[\frac{\tilde{h}_1^{*}\tilde{g}}{S_n}\right](t_c), \label{maxtc}
\end{align}
where $\mathcal{F}^{-1}$ stands for the inverse Fourier transform operator. One can thus numerically perform the FFT on $\tilde{h}_1^{*}\tilde{g}/S_n$ and find the maximum value of $(h|g)$ over $t_c$. Although the result depends on the sampling rate and sometimes the inner product is very sensitive to $t_c$, this FFT method still gives a good initial guess for $t_c$, which can be refined by searching numerically around this trial value (e.g.~through a grid search).

Let us now focus on maximization over the phase offset $l_c$. If $\tilde{h}(f)=\tilde{h}_2(f)e^{i2l_c}$, then
\begin{align}
(h|g)=4 \Re \int\frac{\tilde{h}_2(f)^{*}\tilde{g}(f)}{S_n(f)}e^{-i2l_c}\mathrm{d}f. \label{inner}
\end{align}
Since the mean anomaly at coalescence $l_c$ is a constant, it can be pulled out of the integral, which can be computed to obtain a complex number that we express as $A \, e^{i\delta}$. Equation (\ref{inner}) then becomes 
\begin{align}
(h|g)=4A\cos(\delta-2l_c).
\end{align}
and the inner product can be easily maximized via
\begin{align}
&\max_{l_c} (h|g)=4A =\sqrt{\left(4A\right)^2\left[\cos^2\delta+\cos^2\left(\delta-\frac{\pi}{2}\right)\right]} \notag \\
&=\sqrt{(h_{l_c=0}|g)^2+\left.\left(h_{l_c=\frac{\pi}{4}}\right|g\right)^2}. \label{max-phic}
\end{align}

Clearly, the above maximization procedure is only valid when the waveform model has a single harmonic, so we must now construct a new procedure that works for models with multiple harmonics. Suppose the waveform can be written as
\begin{align}
\tilde{h}=\tilde{h}_2e^{i2l_c}+\sum_{\ell\neq 2}\tilde{h}_\ell e^{i\ell l_c}, \label{multiple-waveform}
\end{align}
and that the overlap between different modes is much smaller than the SNR of each mode, i.e.,
\begin{align}
\left(\tilde{h}_\ell e^{i\ell l_c}|\tilde{h}_k e^{ik l_c}\right)\ll\left(\tilde{h}_\ell e^{i\ell l_c}|\tilde{h}_\ell e^{i\ell l_c}\right), \left(\tilde{h}_k e^{ik l_c}|\tilde{h}_k e^{ik l_c}\right)
\end{align}
with $\ell \neq k$, which is always the case in the PC approximation, but obviously breaks down when the eccentricity is not small. The SNR of $h$ can then be approximated by 
\begin{align}
(h|h)&=\left(\left.\sum_\ell \tilde{h}_\ell(f) e^{i\ell l_c}\right|\sum_\ell \tilde{h}_\ell(f) e^{i\ell l_c}\right) \notag \\
&\sim\sum_\ell \left(\left.\tilde{h}_\ell(f)\right|\tilde{h}_\ell(f)\right),
\end{align}
which does not depend on $t_c$ or $ l_c$. Because of this, the procedure above is still applicable to maximize over $t_{c}$. 

Let us now focus on maximizing the overlap over $ l_c$ in the multiple harmonic case. As before, we pull out the factor of $e^{i\ell l_c}$ and define
\begin{align}
A_\ell e^{i\delta_\ell}&\equiv4\int\frac{\tilde{h}_\ell^*(f)\tilde{g}(f)}{S_n(f)}\mathrm{d}f. \label{define}
\end{align}
such that the inner product between $h$ and $g$ becomes
\begin{align}
&(h|g)=A_2\cos(2 l_c-\delta_2)+\sum_{\ell\neq2} A_\ell\cos(\ell l_c-\delta_\ell)\,.
\end{align}
Defining two new angles
\begin{align}
\psi&\equiv l_c-\frac{\delta_2}{2}, \qquad
\phi_\ell \equiv\frac{\delta_2}{2}\ell-\delta_\ell, \label{phiell}
\end{align}
we can rewrite the inner product as 
\begin{align}
\Delta(\psi)\equiv(h|g)=A_2\cos2\psi+\sum_{\ell\neq2} A_\ell\cos(\ell\psi+\phi_\ell). \label{Delta}
\end{align}
We now maximize $\Delta(\psi)$ using the fact that the amplitude of the $\ell = 2$ mode is usually much larger than that of any other modes in the PC approximation, i.e., $A_2\gg A_\ell$, where we assume $A_{2}\geq 0$ without loss of generality. The maximum angle $\psi_m$ is then near $\psi=0$ and it satisfies $\left.{\mathrm{d}\Delta}/{\mathrm{d}\psi}\right|_{\psi_m}=0$, which reduces to
\begin{align}
0&=\sin2\psi_m+\sum_{\ell\neq2} x_\ell\sin(\ell\psi_m+\phi_\ell),
\end{align}
where we have defined $x_\ell=\ell A_\ell/(2A_2)$. Since $x_\ell$ is a small number for small eccentricity orbits, perturbation theory can be used to solve this equation and find
\begin{align}
\psi_\text{m}=-\frac{1}{2}\sum_{\ell\neq2}  x_\ell \sin\phi_\ell +\mathcal{O}(x_\ell^2).
\end{align}
Substituting this result into Eq. (\ref{Delta}) gives the maximum value of $\Delta$, which reduces to
\begin{align}
\max \Delta&=A_2+\sum_{\ell\neq2} A_\ell\cos\phi_\ell =\sum_\ell A_\ell\cos\phi_\ell, \label{max-phic-multiple}
\end{align}
where we have used that $\phi_2\equiv0$ in the last equality. 

Since $t_c$ does not enter the above maximization over $ l_c$, one could first maximize over the latter by repeating the above calculation with $t_c$ undetermined, which would render $A_\ell$ and $\phi_\ell$ functions of $t_c$. The full maximization procedure then turns into 
\begin{align}
\max_{t_c, l_c}(h|g) &= \max_{t_c}\sum_\ell A_\ell(t_c)\cos\phi_\ell(t_c), \label{over-max-tc-multiple}
\end{align}
which can be carried out numerically. We can however improve the efficiency of the algorithm by choosing a good initial guess, which can be found through the FFT method described above. This method, however, should be slightly modified because of the multiple harmonics in the waveforms. Suppose that the waveform model can be expressed as
\begin{align}
\tilde{h}(f)=e^{-2\pi ift_c}\sum_\ell \tilde{w}_\ell(f) e^{i\ell l_c},
\end{align}
which then implies that
\begin{align}
|(h|g)|&\leq4\Re\left|\sum_\ell e^{-i\ell l_c}\mathcal{F}^{-1}\left[\frac{\tilde{w}_\ell^*\tilde{g}}{S_n}\right](t_c)\right| \notag \\
&\leq 4\sum_\ell\left|\mathcal{F}^{-1}\left[\frac{\tilde{w}_\ell^*\tilde{g}}{S_n}\right](t_c)\right|\equiv \Lambda(t_c).
\end{align}
which defines the new function $\Lambda(t_c)$. Because we have sampled $\tilde{w}_\ell^*$ and $\tilde{g}$ in the frequency domain, the evaluation of $\Lambda(t_{c})$ with a specific sampling rate can be easily achieved through a FFT. The maximum value of the sequence of returned samples provides a good initial guess for $t_c^{(0)}$, because the $\ell = 2$ mode is always much stronger than any other mode for small eccentricities. Then the full maximization can be achieved by numerically evaluating Eq.~(\ref{over-max-tc-multiple}) near $t_c^{(0)}$. We have checked that the maximum point $t_c^{\text{max}}$ is indeed close to $t_c^{(0)}$.

The procedure described above is what we will employ in the next sections to estimate the regime of validity of the PC approximation, which then defines the region inside which we will carry out a Fisher analysis. This procedure is similar to that presented recently in~\cite{Moore:2018kvz}. The main difference is that here we are focused on small eccentricity binaries, and thus, the method described above is tailored made for PC waveforms. The analysis of~\cite{Moore:2018kvz}, on the other hand, is more generic, and valid also for binaries with moderate eccentricities. If one wishes to consider the latter, then the methods of~\cite{Moore:2018kvz} should be employed to maximize over phase and time offsets. 

\begin{table}
    \centering
    \caption{Initial and final frequencies of integration for the two representative sources and different ground-based detectors. We also include the number of points sampled in the time-domain eccentric TaylorT4 model, and the SNRs for the quasi-circular case at a $D_{L} = 100$Mpc.}
    \begin{tabular}{c c c c c c c} \hline\hline
    Sources & Detectors & $f_\text{lo}$(Hz) & $f_\text{hi}$(Hz)& N & SNR & $e_{0}^{\rm{max}}$ \\ \hline
        \multirowcell{10}{BHNS} & \multirowcell{2}{aLIGO}   &\multirowcell{2}{10} &\multirowcell{2}{660.0}  & \multirowcell{2}{$2^{19}$} & \multirowcell{2}{29.6}& \multirowcell{2}{0.15} \\ \\ \cline{2-7}
& \multirowcell{2}{A+} &\multirowcell{2}{10} &\multirowcell{2}{1145.9}& \multirowcell{2}{$2^{19}$} & \multirowcell{2}{42.5}& \multirowcell{2}{0.18} \\ \\ \cline{2-7}
& \multirowcell{2}{Voyager} & \multirowcell{2}{7.2}&\multirowcell{2}{1597.0} & \multirowcell{2}{$2^{21}$} & \multirowcell{2}{140.4}& \multirowcell{2}{0.20} \\ \\ \cline{2-7}
& \multirowcell{2}{CE} & \multirowcell{2}{5.3} &\multirowcell{2}{1928.6} & \multirowcell{2}{$2^{23}$}& \multirowcell{2}{693.7}&  \multirowcell{2}{0.17}\\ \\ \cline{2-7}
& \multirowcell{2}{ET-D} & \multirowcell{2}{1.5} &\multirowcell{2}{1928.6}& \multirowcell{2}{$2^{28}$} & \multirowcell{2}{431.2}& \multirowcell{2}{0.22}  \\ \\ \hline
        \multirowcell{10}{BNS} & \multirowcell{2}{aLIGO}   &\multirowcell{2}{10.2}  & \multirowcell{2}{398.5} & \multirowcell{2}{$2^{20}$} & \multirowcell{2}{14.5}& \multirowcell{2}{0.18}\\ \\ \cline{2-7}
& \multirowcell{2}{A+} &\multirowcell{2}{10.2} & \multirowcell{2}{703.3} & \multirowcell{2}{$2^{21}$}& \multirowcell{2}{21.5}& \multirowcell{2}{0.21} \\ \\ \cline{2-7}
& \multirowcell{2}{Voyager} & \multirowcell{2}{7.6} & \multirowcell{2}{1064.0} & \multirowcell{2}{$2^{22}$} & \multirowcell{2}{69.8}& \multirowcell{2}{0.18} \\ \\ \cline{2-7}
& \multirowcell{2}{CE} & \multirowcell{2}{5.5} & \multirowcell{2}{1899.5} & \multirowcell{2}{$2^{25}$} &\multirowcell{2}{342.7} & \multirowcell{2}{0.14}\\ \\ \cline{2-7}
&\multirowcell{2}{ET-D} & \multirowcell{2}{1.6}& \multirowcell{2}{1423.8} & \multirowcell{2}{$2^{29}$} & \multirowcell{2}{213.4}& \multirowcell{2}{0.14}  \\
 \\ \hline\hline
     \end{tabular}
\label{table:overlap}
\end{table}
\subsection{Validity of the PC model \\ for ground-based detectors}
\label{sec:overlap-gound}

\begin{figure*}[htb]
        \includegraphics[width=\columnwidth,clip=true]{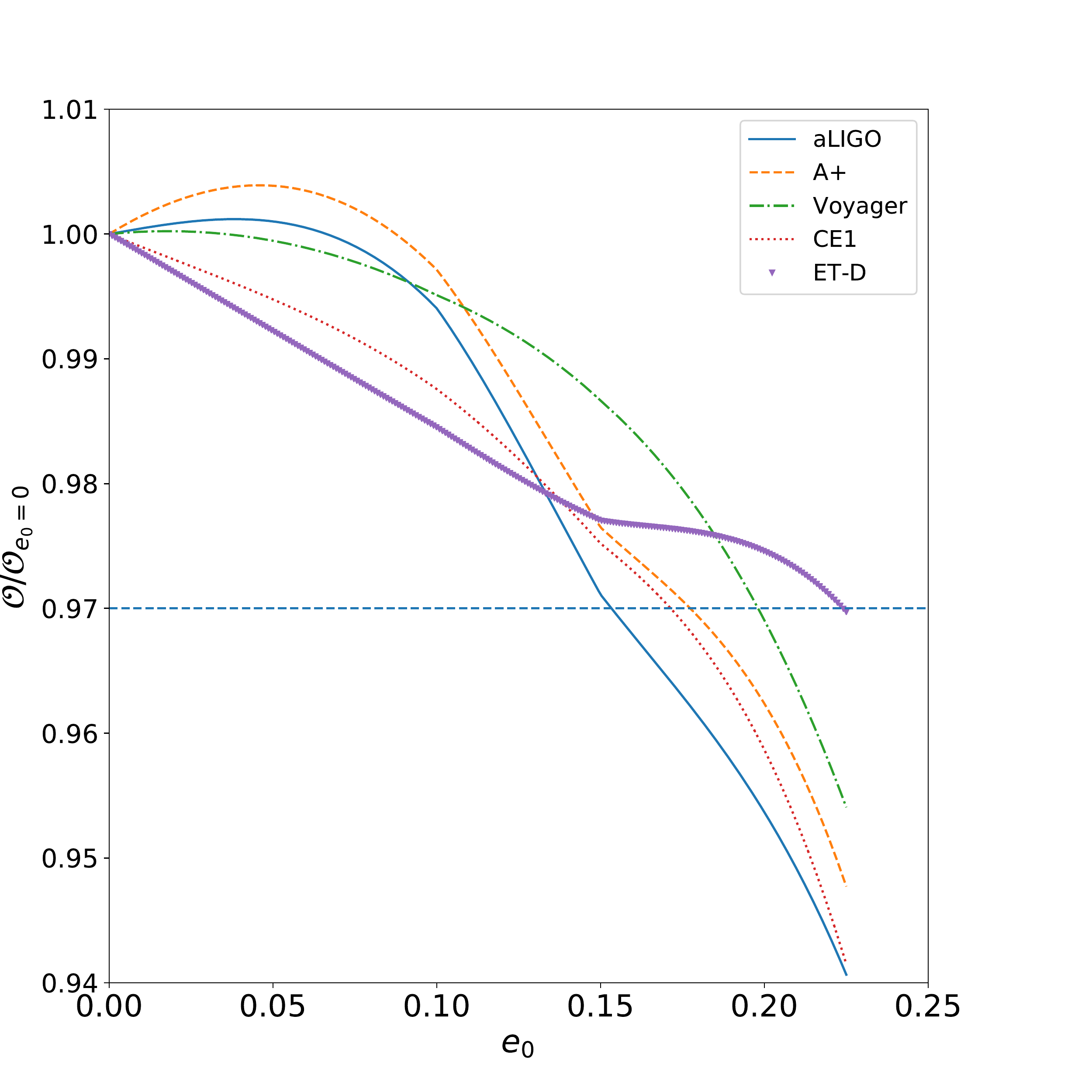}
        \includegraphics[width=\columnwidth,clip=true]{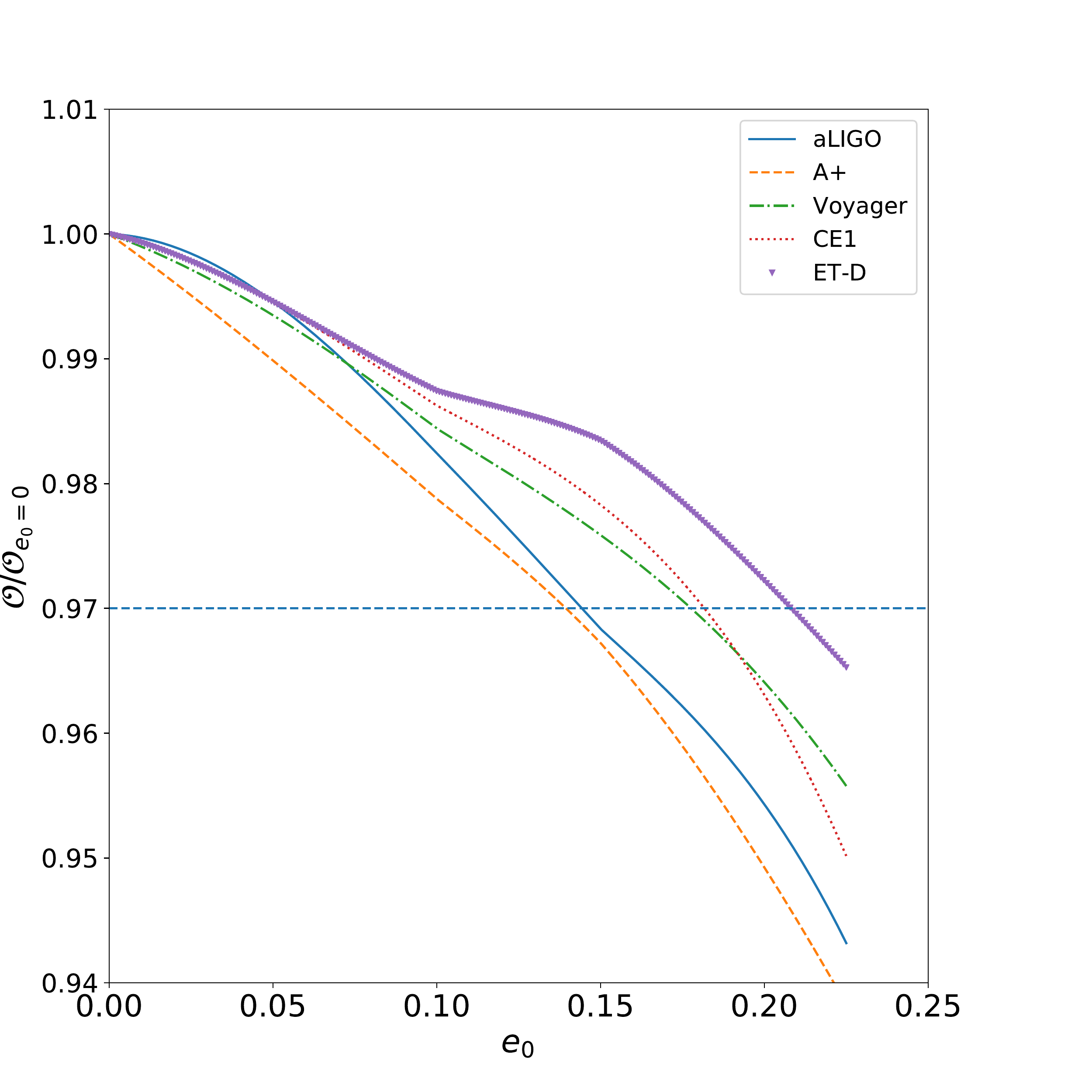}
  \caption{(Color Online) The normalized match for ground-based detectors as a function of the initial eccentricity $e_0$ for the representative BHNS (left) and the BNS (right) . The dashed line corresponds to the 0.97 threshold. Observe that the PC is valid up to initial eccentricities around $0.15$--$0.25$ depending on the detector.}
 \label{fig:overlap-ground}
\end{figure*}

We here search for an estimate of the maximum initial eccentricity that our analytic model can tolerate by calculating the match between it and the eccentric TaylorT4 model in the GR limit. The maximization over $(t_{c}, l_c)$ is carried out as explained in the previous subsection. For simplicity, we work in the sky-averaged approximation, i.e.~averaging over all angular parameters, such as $\theta_S$, $\phi_S$, $\psi_S$, $\iota$ and $\beta$, and we here focus on ground-based detectors (with LISA discussed in the next subsection).

Two representative sources are considered: (i) a BH-NS binary with component masses $(10M_\odot,1.4M_\odot)$, and (ii) a neutron star binary (BNS) with masses $(1.2M_\odot,1.8M_\odot)$, both at a fixed luminosity distance $D_L$ of 100Mpc; we list the SNRs in the quasi-circular case for each detector in Table~\ref{table:overlap}. We do not consider BH binaries because scalar radiation is suppressed in vacuum by the no-hair theorems. We ignore spins all together in this paper, as this is beyond the scope of this paper.

All overlap calculations require the specification of a starting and ending frequency of integration, $f_\text{lo}$ and $f_{\text{hi}}$ respectively. We here choose 
\begin{align}
f_\text{lo}^\text{grnd} &=f_\text{lratio} \qquad
f_\text{hi}^\text{grnd}=\min(f_\text{hratio},10F_\text{ISCO}),
\end{align}
where $f_\text{lratio}$ and $f_\text{hratio}$ are the low and the high frequencies at which the amplitude of the GW model is 10$\%$ of the spectral noise. The absolute maximum of  $10F_\text{ISCO}$ stems from the SPA condition $f=\ell F$, the fact that we keep ten harmonics in the waveforms (so the highest GW frequency that the system can emit is $10F_\text{ISCO}$), and the need to ensure the PN approximation does not break down. We list the initial and final GW frequencies of integration, as well as the number of points sampled in the numerical model for different detectors, in Table. \ref{table:overlap}.

Figure \ref{fig:overlap-ground} shows the match as a function of the initial eccentricity $e_0$ for both the representative BHNS and BNS systems discussed above, normalized to the match in the quasi-circular case. Comparing the two figures, we see that the match clearly depends on the source and the detector modeled. In the BHNS case, the normalized match computed with second-generation detectors increases slightly in the small initial eccentricity region, which simply means that the eccentric match is slightly larger than the quasi-circular one; we have checked that all of the matches computed are smaller than unity, as expected from the Cauchy-Schwarz inequality. For third generation detectors, and when we consider the BNS case, the match decreases monotonically with initial eccentricity. Observe also that the PC model is accurate up to eccentricities of roughly $0.14$--$0.22$, at which point the match drops below the 97\% threshold. The last column of Table \ref{table:overlap} shows the initial eccentricities at which the match intersects the threshold. 


\subsection{LISA}
\label{sec:overlap-lisa}

Let us now focus on the validity of the PC approximation for LISA sources. As in the ground-based case, we will work with the sky-averaged match, with the maximization over $(t_{c}, l_c)$ carried out as explained earlier in this section. For LISA, however, we will assume a 5 year mission duration and consecutive observation, and we will focus on BHNS binaries only (as BNSs have too low an SNR in the LISA band), with a representative system composed of compact objects with masses $(10^{2},1.4) M_{\odot}$. We further place the binary at a luminosity distance $D_L$ of 20Mpc to ensure the SNR is large enough for the signal to be detectable. We could have picked a BH with a much larger mass to make the signal more easily detectable, but if we had done so, we would have entered into a parameter region in which the PN approximation becomes highly inaccurate. Indeed, the loss of accuracy of the PN approximation in the EMRI limit has been studied in some detail in the past~\cite{Yunes:2008tw,Yunes:2009ef}. In Appendix~\ref{sec:app-overlap}, we show how the match deteriorates monotonically with mass ratio $q$ in the quasi-circular case; if q is smaller than $7\times10^{-3}$, which corresponds to a binary with component masses of roughly $(200 M_\odot,1.4M_\odot)$, the overlap becomes smaller than 0.97. Similar results were recently reported in~\cite{Moore:2019xkm}.

As in the last subsection, we choose different integration limits for the overlap based on the noise curve studied. In particular, we choose
\begin{align}
f_\text{lo}^\text{LISA}&=\max(f_\text{lratio},f_\text{5 years}), 
\\
f_\text{hi}^{\text{LISA}} &=\min(f_\text{hratio},10F_\text{ISCO}).
\end{align}
where $f_\text{5 years}$ is the GW frequency 5 years before merger. For the system we considered, this means $f_\text{lo}^\text{LISA}=35.6$ mHz and $f_\text{hi}^\text{LISA}=473.0$ mHz

\begin{figure}
\centering
\includegraphics[width=\columnwidth,clip=true]{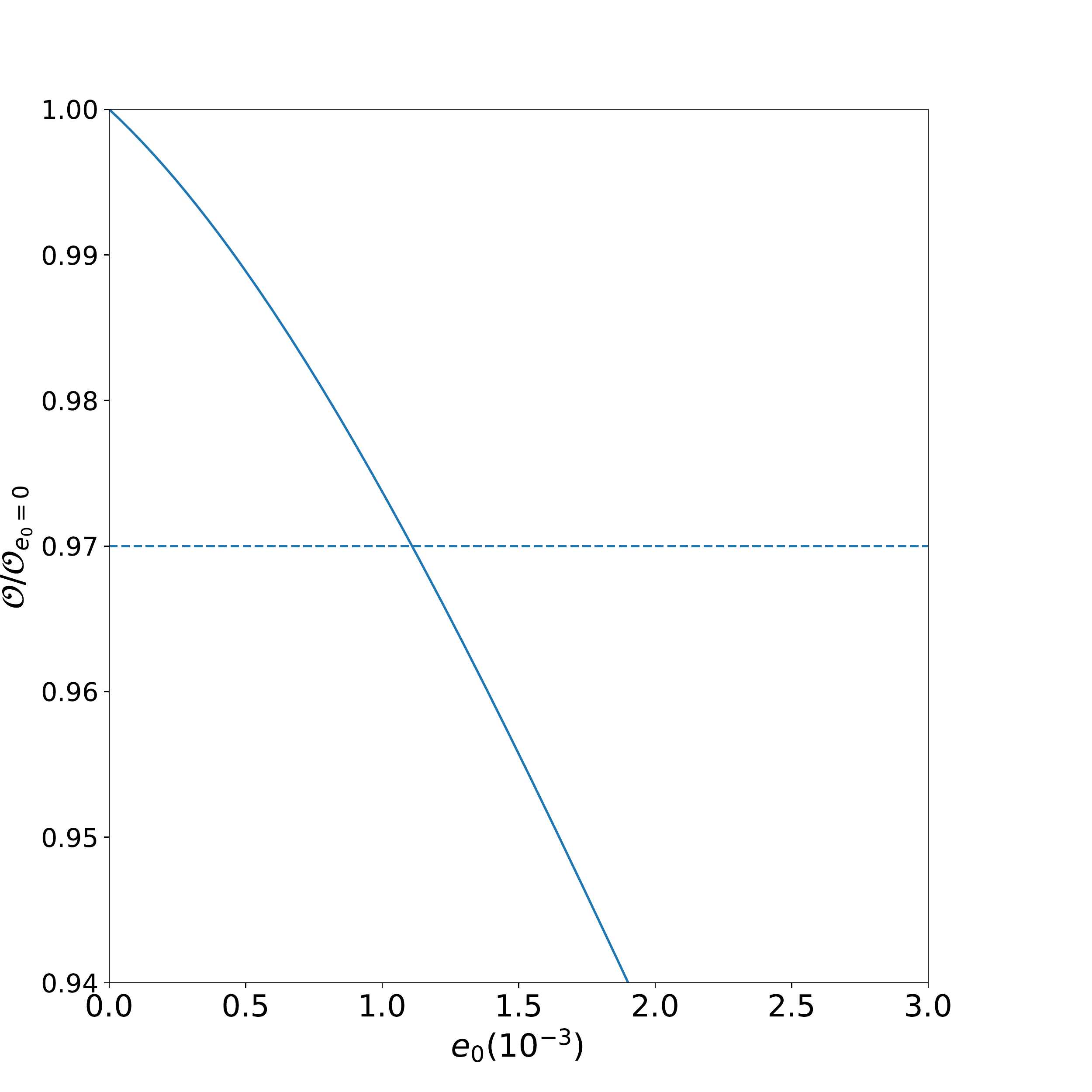}
\caption{(Color Online) The normalized overlap for LISA as the function of $e_0$. GWs are emitted by BHNS with $(100M_\odot,1.4M_\odot)$. The dashed line is the threshold.}
\label{fig:overlaplisa}
\end{figure}

Figure \ref{fig:overlaplisa} shows the normalized match as a function of the initial eccentricity $e_0$. As compared to ground-based detectors, the match deteriorates much more rapidly with initial eccentricity. The intersection of the match with a 97\% threshold yields the maximum eccentricity $e_0^{\rm{max}} \approx 10^{-3}$. We we will see in the next subsection, this is a very small regime of validity in initial eccentricity, which begs for the further development of accurate intermediate mass-ratio inspiral models both in the quasi-circular and eccentric case. 

\section{Eccentricity effect on constrains of $\omega$}
\label{sec:fisher-matrix}

In this section, we carry out a sky-averaged, Fisher analysis to discuss the effect of eccentricity on constraints on $\omega$. That is, we assume that we have detected a GW consistent with GR and we estimate the accuracy to which we can state that the $b$ parameter of Jordan-Brans-Dicke-Fierz theory is statistically consistent with zero. We begin by reviewing the basics of a Fisher analysis, and we then present results for both ground- and space-based detectors.

\subsection{The basics of a Fisher analysis}
\label{sec:basics-fisher}
Suppose the measured data $s(t)$ consists of a signal $h(t)$ and random noise $n(t)$, i.e., 
\begin{align}
s(t)=h(t)+n(t).
\end{align}
If the detector noise is stationary and Gaussian, then the likelihood function is
\begin{align}
\label{likelihood}
p(s|\bm{\theta})\propto e^{-(s-h|s-h)/2},
\end{align}
where $\bm{\theta}$ is the model parameter vector and the inner product $(s-h|s-h)$ was defined in Eq. (\ref{innerproduct}). When the SNR is large, the likelihood function in Eq. (\ref{likelihood}) can be approximated by 
\begin{align}
p(s|\bm{\theta})\propto e^{-\Gamma_{mn}\Delta\theta^m\Delta\theta^n/2},
\end{align}
where the Fisher information matrix, $\Gamma_{mn}$, is given by
\begin{align}
\Gamma_{mn}=\left(\left.\frac{\partial h}{\partial \theta^m}\right|\frac{\partial h}{\partial \theta^n}\right).
\end{align}
In our case, the Fisher matrix is 7-dimensional because the model parameters are  $\bm{\theta} = [ l_c,t_c,\ln D_L, b, \eta,\ln \mathcal{M}, e_0]$.

The Fisher matrix sets a lower bound, i.e., the Cramer-Rao bound, for the statistical covariance of estimated parameters, namely 
\begin{align}
\text{covar}(\theta^m,\theta^n)\geq(\Gamma^{-1})_{mn}.
\end{align}
Equality holds in the high SNR or linearized-signal approximation \citep{Vallisneri08}. In our paper, we only work in the high SNR limit, so that the Fisher matrix is a good quadratic approximation to the peak of the likelihood function. The diagonal components of the covariance matrix return the variance of a measured parameter, namely 
\be
\sigma_m=\sqrt{(\Gamma^{-1})_{mm}}, 
\ee
which in our case provides an estimate for the $1$--$\sigma$ upper bound on $b$. From this upper bound, a lower bound on $\omega$ can be obtained through Eq. (\ref{b-s}). The sensitivity difference $\mathcal{S}$ is calculated based on the APR equation of state and Eq. (\ref{sensitivity-approximation}), as discussed in Sec. \ref{sec:orbits}.

\subsection{Ground-based detectors}
\label{sec:fisher-ground}

Let us now consider the effect of eccentricity in projected constraints on the $\omega$ coupling parameter of Jordan-Brans-Dicke-Fierz theory with ground-based detectors. As before, we consider two representative sources: (i) a BHNS binary with component masses $(10M_\odot,1.4M_\odot)$, and (ii) a BNS with masses $(1.2M_\odot,1.8M_\odot)$. The luminosity distance $D_L$ is chosen again to be 100Mpc, with the SNR, $f_\text{lo}$ and $f_\text{hi}$ given in Table \ref{table:overlap}. The NS sensitivities for objects with mass 1.2, 1.4, 1.8$M_\odot$ are 0.140, 0.171 and 0.245, respectively. 

\begin{figure}[htb]
        \includegraphics[width=8.6cm,clip=true]{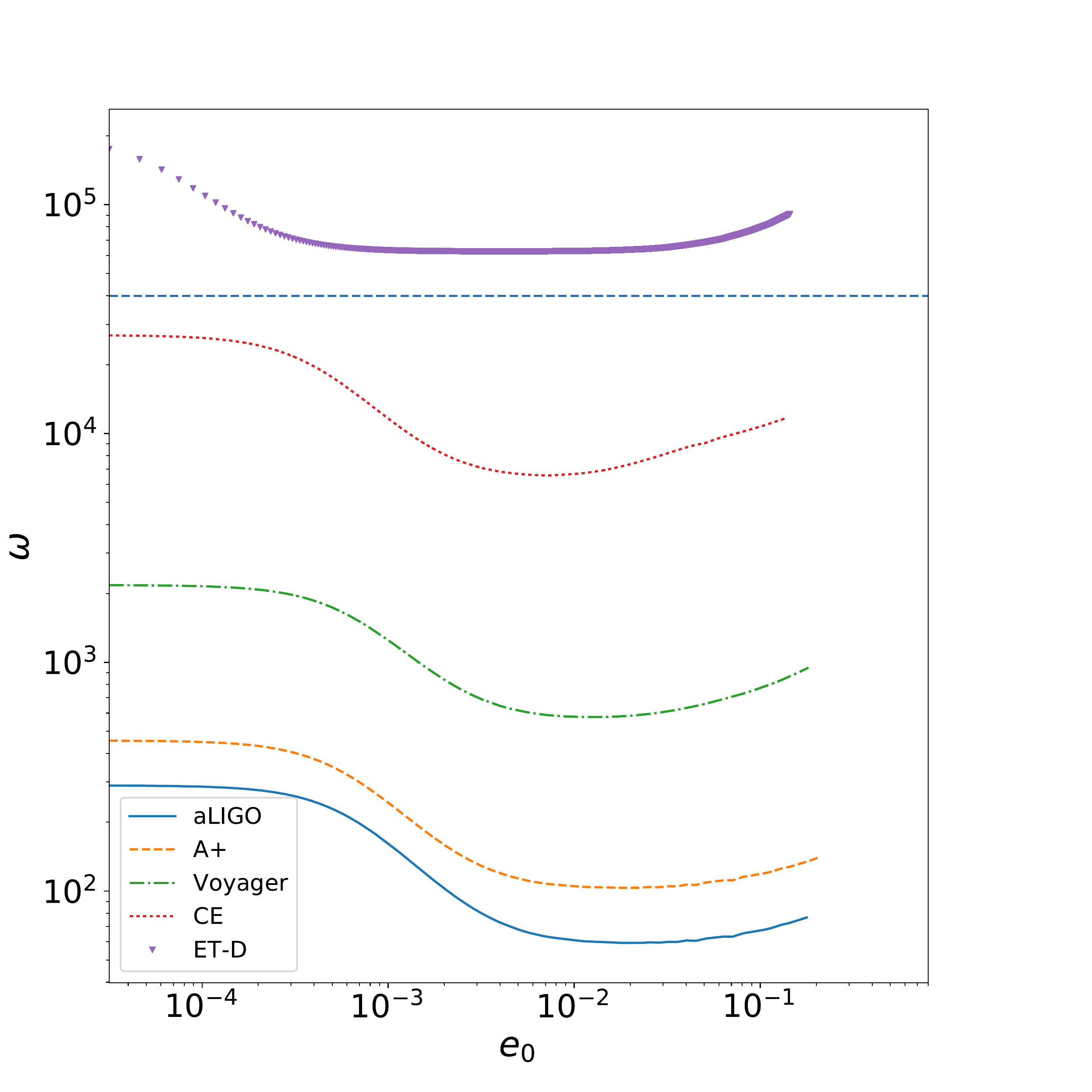}
  \caption{(Color Online) Projected constrains or lower bounds on $\omega$ as a function of initial eccentricity for a BNS signal using current and future ground-based detectors. All curves are terminated at the maximum initial eccentricity listed in Table \ref{table:overlap} and the horizon dashed line is the current constraint on $\omega$ from tracking of the Cassini spacecraft~\cite{Bertotti03}. Once the initial eccentricity is large enough, the projected constraint on $\omega$ is enhanced by eccentricity.}
 \label{fig:omega-ground-NSNS}
\end{figure}
Figures~\ref{fig:omega-ground} and~\ref{fig:omega-ground-NSNS} show projected constraints (lower bounds) on $\omega$ as the function of the initial eccentricity $e_0$, terminating all curves at the maximum $e_0$ found in Table \ref{table:overlap}. When $e_{0}$ is very small, the projected constraints we obtain are consistent with those found in the quasi-circular limit. Observe that the constraint improves with detector upgrade mostly because we fix the luminosity distance, which implies the SNR increases as the noise decreases. In the ET case, the constraint improves because the signal can be sampled at a lower starting frequency than in the CE case, which enhances modified gravity effects that enter at negative PN orders. We also see that the only way to beat current Solar System constraints (the horizontal dashed line) is to either use third-generation detectors or to go to higher $e_{0}$. 

\begin{figure}
\centering
\includegraphics[width=\columnwidth,clip=true]{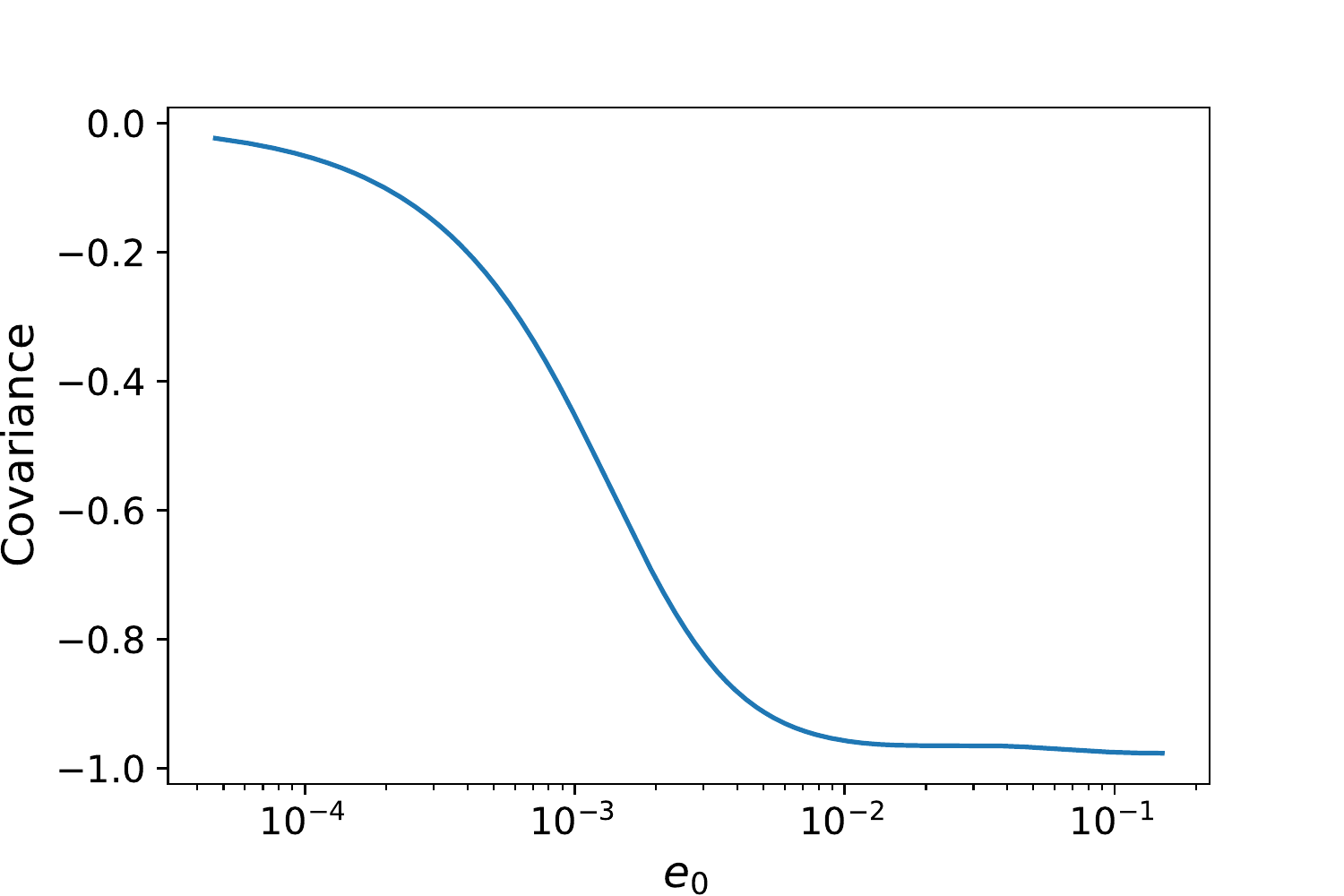}
\caption{(Color Online) The covariance of $b$ and $e_0$ as the function of $e_0$. The detector is CE and the source is BHNS inspiral. The covariance peaks around $e_0\sim 0.009$, which is also the minimum point of $e_0-\omega$ curves.}
\label{fig:ecc-cov}
\end{figure}

Figures~\ref{fig:omega-ground} and~\ref{fig:omega-ground-NSNS} also show the effect of the initial eccentricity on the $\omega$ constraint. First, observe that when the eccentricity is small (e.g.~when it is smaller than $10^{-2}$), the projected constraint deteriorates relative to the quasi-circular projection. This is because a partial degeneracy between the $e_{0}$ and $b$ parameters in the Fisher matrix emerges in this regime. Figure \ref{fig:ecc-cov} shows the covariance between $e_0$ and $b$ as a function of $e_0$, using the CE detector and the BHNS binary source as a representative example. Observe that an anti-correlation emerges between $e_{0}$ and $b$ precisely in the eccentricity regime inside which the projected constraints on $\omega$ in Fig.~\ref{fig:omega-ground} also deteriorate. However, once the initial eccentricities is above $0.01$, the projected constraints begin to improve, as summarized in Table \ref{table:omega-ground-bhns}.


\begin{table}
    \centering
    \caption{Projected constraints (lower bounds) on $\omega$ using GWs from BHNS (top) and BNS (bottom) inspirals with different initial eccentricity and observed with different detectors. We here list three situations: (i) constraints in the quasi-circular case ($e_0=0$), (ii) the worst constraint on $\omega$ and (iii) the constraint evaluated at the maximum initial eccentricity of Table \ref{table:overlap}. We also list the corresponding suppression factors relative to the quasi-circular constraint.}
    \begin{tabular}{l c c c c c c} \hline\hline
  \multirow{8}{*}{BHNS} & Detectors & aLIGO & A+ & Voyager & CE & ET-D \\ \cline{2-7}
 &Circular & 600 & 779 & 3,577 & 44,473 & 360,661\\ \cline{2-7}
 &Worst & 135 & 183 & 1,027 & 12,378 & 112,734 \\ \cline{2-7}
 &$e_{0}^{\rm{max}}$ & 161 & 229 & 1,691 & 21,523 & 222,364\\ \cline{2-7}
 &Sup.~Factor  & \multirow{2}{*}{0.27}  & \multirow{2}{*}{0.29} & \multirow{2}{*}{0.47} & \multirow{2}{*}{0.48} & \multirow{2}{*}{0.62} \\ 
 &($e_{0}^{\rm{max}}$) & & & & & \\ \cline{2-7}
 &Sup.~Factor & \multirow{2}{*}{0.23} & \multirow{2}{*}{0.23} & \multirow{2}{*}{0.29} & \multirow{2}{*}{0.28} & \multirow{2}{*}{0.31}  \\ 
 &(Worst) & & & & & \\  \hline
  \multirow{8}{*}{BNS} & Detectors & aLIGO & A+ & Voyager & CE & ET-D \\ \cline{2-7}
 & Circular & 289 & 454 & 2,174 & 26,848 & 195,766\\ \cline{2-7}
 & Worst & 59 & 103 & 576 & 6,560 & 62,582 \\ \cline{2-7}
 & $e_{0}^{\rm{max}}$  & 77 & 141 & 947 & 11,731 & 91,095\\ \cline{2-7}
 & Sup.~Factor & \multirow{2}{*}{0.27} & \multirow{2}{*}{0.31} & \multirow{2}{*}{0.44} & \multirow{2}{*}{0.44} & \multirow{2}{*}{0.47} \\ 
  &($e_{0}^{\rm{max}}$) & & & & & \\ \cline{2-7}
 & Sup.~Factor & \multirow{2}{*}{0.20} & \multirow{2}{*}{0.23} & \multirow{2}{*}{0.26} & \multirow{2}{*}{0.24} & \multirow{2}{*}{0.32}  \\ 
 &(Worst) & & & & & \\  \hline\hline
     \end{tabular}
\label{table:omega-ground-bhns}
\end{table}

\begin{figure}[htb]
        \includegraphics[width=\columnwidth,clip=true]{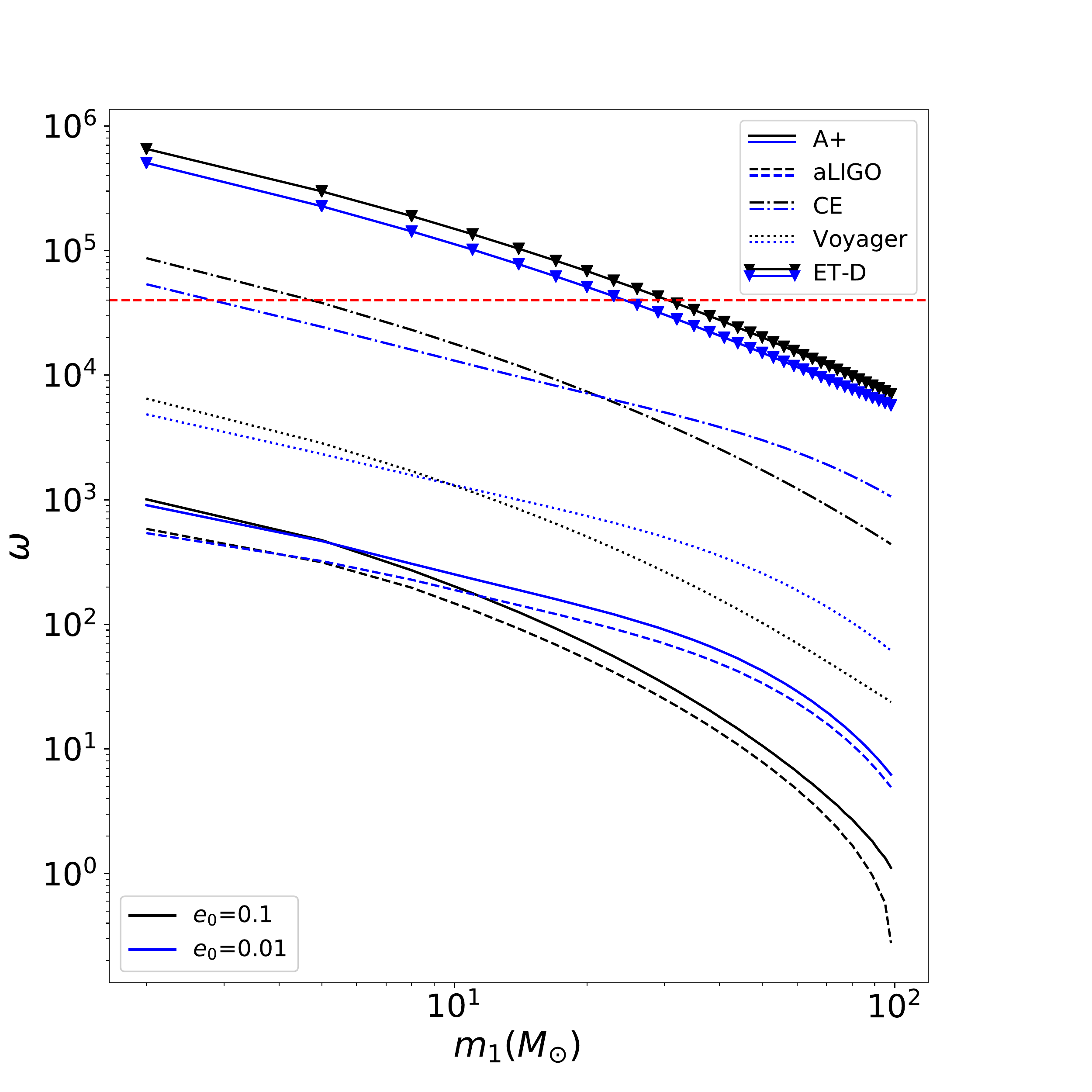}
 \caption{ (Color Online) Projected constraints on $\omega$ as a function of BH mass in BHNS inspirals with initial eccentricities $0.01$ (blue) and $0.1$ (black). The horizontal dashed line corresponds to the current constraint on $\omega$ from the tracking of the Cassini spacecraft~\cite{Bertotti03}. Observe that the constraint improves as the BH mass decreases.}
 \label{fig:mass-ground}
\end{figure}
We conclude this analysis with a short investigation of how the projected constraints scale with total mass of the source. We focus on BHNS inspirals only, with the mass of the NS fixed at $1.4M_\odot$ and the BH mass allowed to vary from $2M_\odot$ to 100$M_\odot$. Figure~\ref{fig:mass-ground} shows projected constraints as a function of the BH mass for systems with initial eccentricity $e_0=0.01$ and $e_0=0.1$. Observe that the projected constraints on $\omega$ deteriorates monotonically with increasing BH mass, which is consistent with the results of~\citep{Will94} in the quasi-circular limit. This is because the higher the BH mass, the shorter the inspiral signal in the detector band, since we do not consider here the merger phase of coalescence.

\subsection{Space-based Detectors}
\label{sec:fisher-lisa}
Let us now consider projected constraints on $\omega$ using LISA. As discussed in Sec. \ref{sec:overlap-lisa}, we study BHNS inspirals with component masses $(100M_\odot,1.4M_\odot)$ to avoid inaccuracies in the PN approximation for intermediate mass-ratio inspirals. The luminosity distance $D_L$ is still kept at 20Mpc and we continue to consider a 5 year long observation to ensure the signal is detectable. With its three arms, LISA represents a pair of two orthogonal arm detectors, I and II, producing two linearly independent signals. The relation between pattern functions of the two detectors can be expressed as \citep{Mikoczi12}
\begin{align}
F^{II}_{+,\times}=F^{I}_{+,\times}\left(\phi_S-\frac{\pi}{4}\right).
\end{align}
But since our Fisher analysis is sky-averaged, the signals in the two detectors can be treated as identical, and so we assume the two interferometers detect the signals simultaneously, which simply leads to an SNR enhancement of $\sqrt{2}$. In addition, because of the triangular shape of the LISA configuration, the strain in Eq. (\ref{pattern}) must be rescaled as $h=\sqrt{3}/2(F_+h_++F_\times h_\times)$.

\begin{figure}[htb]
\centering
\includegraphics[width=\columnwidth,clip=true]{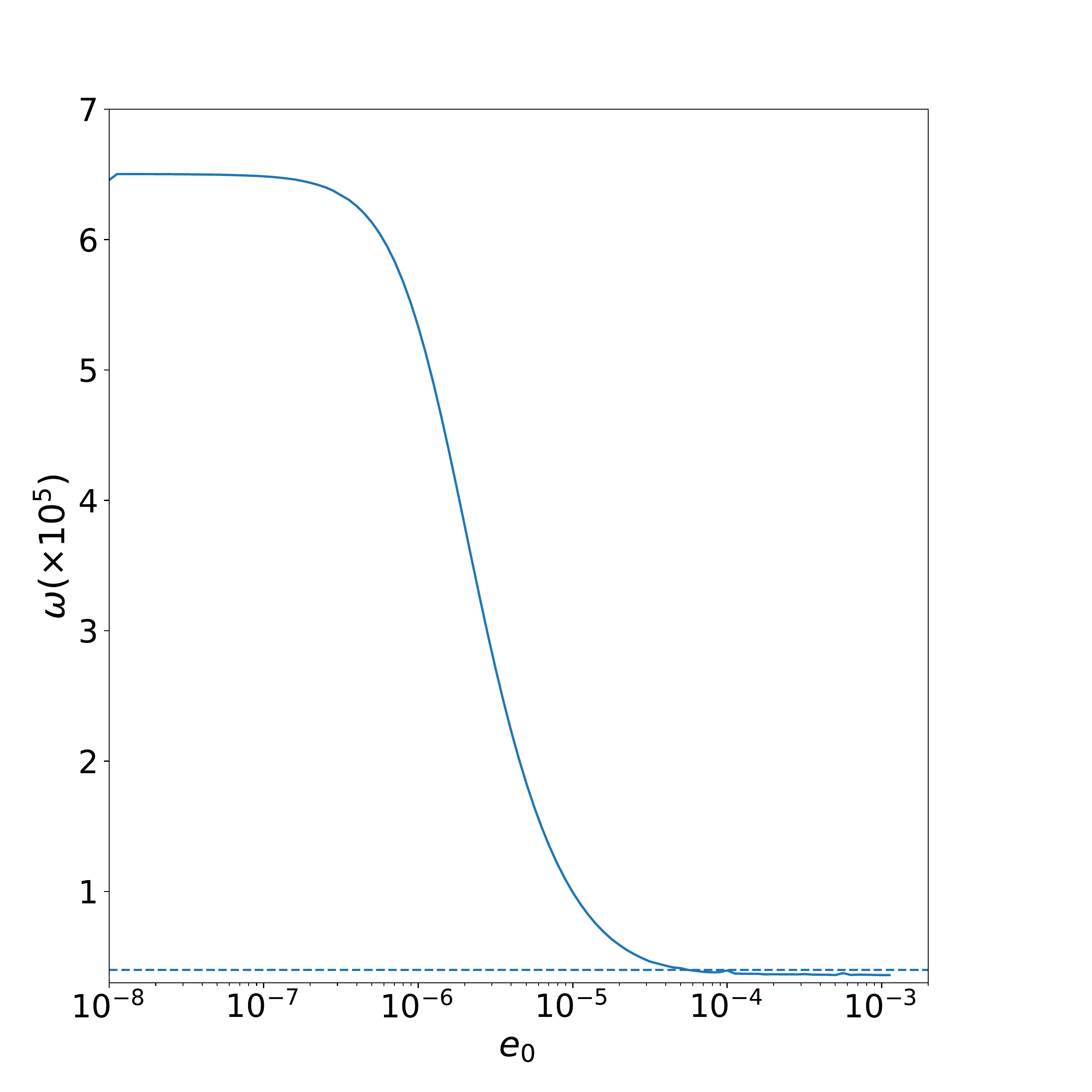}
\caption{(Color Online) Projected constraints on $\omega$ as the function of initial eccentricity $e_0$ for a BHNS signal with component masses $(100M_\odot,1.4M_\odot)$ detected by LISA. The curve is terminated at the maximum initial eccentricity allowed by the PC model that we found in Sec. \ref{sec:overlap-lisa}. The horizontal dashed line represents the current constrain on $\omega$ from the tracking of the Cassini spacecraft~\cite{Bertotti03}.}
\label{fig:lisa}
\end{figure}
Figure~\ref{fig:lisa} shows the projected constraint on $\omega$ as a function of initial eccentricity $e_0$, terminating the curve at the maximum initial eccentricity we found in Sec. \ref{sec:overlap-lisa}. As in the ground-based case, the projected constraints deteriorate initially with increasing eccentricity due to covariances between the $e_{0}$ and the $b$ parameters. In this case, however, the regime of validity of the PC model is so small that we are not able to study larger initial eccentricities that show the turn around and recovery of the projected constraint on $\omega$. In spite of this, the constraints obtained with LISA are the best of all instruments considered, except maybe for third-generation ground-based detectors that could lead to comparable constraints. 

\section{conclusions and discussion}
\label{sec:conclusions}

We have studied the effect of eccentricity in tests of GR with GW observations, focusing on Jordan-Brans-Dicke-Fierz theory as an example and a good initial training step. We began by constructing an analytic, Fourier-domain gravitational waveform for eccentric inspirals in this theory in the PC, PN and SPA approximations. We then estimated the maximum initial eccentricity that can be tolerated by this analytic model by computing the match between it and a 3PN eccentric TaylorT4 model in the GR limit. As a byproduct of this analysis, we also developed a technique to analytically maximize the overlap over a constant phase and time offset when dealing with waveforms composed of multiple harmonics, provided one of them is dominant (i.e.~in the PC limit). We concluded with a Fisher analysis that estimated the accuracy to which the Jordan-Brans-Dicke-Fierz coupling parameter $\omega$ can be constrained with future GW observations consistent with GR using current and third-generation, ground- and space-based detectors. 

We found a variety of interesting results. First, the validity of the PC model is limited to comparable-mass systems with eccentricities smaller than $0.1$ for ground-based detectors, and smaller than $10^{-3}$ for unequal-mass binaries with space-based detectors. Second, constraints on $\omega$ deteriorate as the eccentricity is increased, even in the PC regime, due to partial degeneracies between $\omega$ and the eccentricity parameter of the waveform model. Eventually, as the initial eccentricity is increased, this degeneracy begins to break, and the projected constraints recover, possibly leading to an enhancement for moderately eccentric signals. 

Our results indicate that the correct inclusion of eccentricity in modified gravity GW models is crucial to extract the most information from future signals. Future work could be focused on the development of modified gravity waveforms for systems with moderate eccentricity, for example following the work of~\cite{Moore:2018kvz,Moore:2019xkm}. Such an analysis could confirm whether the projected constraints truly do recover for such much more eccentric signals. Another possible avenue for future work is to consider the inclusion of eccentricity in other modified theories, such as in dynamical Chern-Simons gravity and Einstein-dilaton-Gauss-Bonnet gravity~\cite{Yunes:2013dva}.  Another important issue is to develop eccentric and quasi-circular waveforms for intermediate mass-ratio inspirals even within GR. We have found that for mass ratios more extreme than $1:100$ the overlap between numerical and analytical PN models in GR deteriorates rapidly with mass ratio, specially when using highly-sensitive third-generation detectors, a result also recently reported in~\cite{Moore:2019xkm}. Without accurate models in GR for such systems, it will be difficult to carry out precision tests of Einstein's theory with space-based instruments.  

\acknowledgements 
N.Y acknowledges support from the NSF CAREER grant PHY-1250636 and NASA grants NNX16AB98G and 80NSSC17M0041. We would also like to thank Nicholas Loutrel, Alejandro C\'ardenas-Avenda\~no, Hector Okada da Silva, Blake Moore, and Travis Robson for many discussions.  The research carried out here was partially done on the Hyalite High Performance Computing Cluster of Montana State University, as well as on the Caltech High Performance Cluster, partially supported by a grant from the Gordon and Betty Moore Foundation.
\appendix
\section{The $\varpi_\ell$ Coefficients}
\label{sec:app-xi}

\begin{widetext}
\allowdisplaybreaks[4]
We expand $\varpi_\ell$ to $\mathcal{O}(e^8)$ and $\mathcal{O}(b^1)$, and express $\varpi_\ell$ as $\varpi^{{\GR}}_\ell+\varpi^{{\BD}}_\ell$. The GR part can be found in Appendix C of Ref. \citep{Yunes09}. Below we list the different $\varpi^{{\BD}}_\ell$.

\begin{align}
&\varpi^{{\BD}}_1=b\eta^{2/5}\left(\frac{2\pi f\mathcal{M}}{\ell}\right)^{-2/3}\left[\frac{3}{2}(F_++iF_\times)e 
-\left(\frac{1073}{96}iF_\times+\frac{1045}{96}F_+\right)e^3+\left(\frac{149155}{3072}i F_\times+\frac{47809}{1024}F_+\right)e^5 
\right. \nn \\ 
&\left. -\left(\frac{123518591}{737280}F_++\frac{42997153}{245760}iF_\times\right)e^7\right], \notag \\
&\varpi^{{\BD}}_2=b\eta^{2/5}\left(\frac{2\pi f\mathcal{M}}{\ell}\right)^{-2/3}\left[-2(iF_\times+F_+)
+\frac{149}{8}(iF_\times+F_+)e^2-\left(\frac{71381}{768}F_++\frac{71573}{768}iF_\times\right)e^4 
\right. \nn \\
&\left.
+\left(\frac{197711179}{552960}F_++\frac{198602251}{552960}iF_\times\right)e^6 
+\left(-\frac{43751145037}{35389440}F_+-\frac{43979478733}{35389440}iF_\times\right)e^8\right], \notag \\
&\varpi^{{\BD}}_3=b\eta^{2/5}\left(\frac{2\pi f\mathcal{M}}{\ell}\right)^{-2/3}\left[-\frac{9}{2}(iF_\times+F_+)e
+\frac{1323}{32}(iF_\times+F_+)e^3-\left(\frac{1057323}{5120}F_++\frac{1058547}{5120}iF_\times\right)e^5 
\right. \nn \\
&\left.
+\left(\frac{196117861}{245760}F_++\frac{196502989}{245760}iF_\times\right)e^7\right], \notag \\
&\varpi^{{\BD}}_4=b\eta^{2/5}\left(\frac{2\pi f\mathcal{M}}{\ell}\right)^{-2/3}\left[-8(iF_\times+F_+)e^2
+\frac{149}{2}(iF_\times+F_+)e^4-\left(\frac{1087163}{2880}F_++\frac{1087867}{2880}iF_\times\right)e^6 
\right. \nn \\
&\left.
+\left(\frac{285208745}{193536}F_++\frac{1427629069}{967680}iF_\times\right)e^8\right], \notag \\
&\varpi^{{\BD}}_5=b\eta^{2/5}\left(\frac{2\pi f\mathcal{M}}{\ell}\right)^{-2/3}\left[-\frac{628}{48}\left(iF_\times+F_+\right)e^3
+\frac{11875}{96}\left(iF_\times+F_+\right)e^5-\frac{329260625}{516096}\left(F_++iF_\times\right)e^7\right], \notag \\
&\varpi^{{\BD}}_6=b\eta^{2/5}\left(\frac{2\pi f\mathcal{M}}{\ell}\right)^{-2/3} 
\left[-\frac{81}{4}(iF_\times+F_+)e^4+\frac{62937}{320}(iF_\times+F_+)e^6 
-\left(\frac{14820705}{14336}F_++\frac{74124261}{71680}iF_\times\right)e^8\right], \notag \\
&\varpi^{{\BD}}_7=b\eta^{2/5}\left(\frac{2\pi f\mathcal{M}}{\ell}\right)^{-2/3}\left[-\frac{117649}{3840}(iF_\times+F_+)e^5
+\frac{6235397}{20480}(iF_\times+F_+)e^7\right], \notag \\
&\varpi^{{\BD}}_8=b\eta^{2/5}\left(\frac{2\pi f\mathcal{M}}{\ell}\right)^{-2/3}\left[-\frac{2048}{45}(iF_\times+F_+)e^6
+\frac{145792}{315}(iF_\times+F_+)e^8\right], \notag \\
&\varpi^{{\BD}}_9=-b\eta^{2/5}\left(\frac{2\pi f\mathcal{M}}{\ell}\right)^{-2/3}\frac{4782969}{71680}(iF_\times+F_+)e^7, \notag \\
&\varpi^{{\BD}}_{10}=-b\eta^{2/5}\left(\frac{2\pi f\mathcal{M}}{\ell}\right)^{-2/3}\frac{390625}{4032}(iF_\times+F_+)e^8.
\end{align}

\section{The 3PN waveform phase in GR within the PC approximation}
\label{sec:phase-GR-PN}

We here present the waveform phase in GR $\Psi^{(\ell)}_{\text{GR}}$ to 3PN order and with as many eccentricity corrections as calculated in the literature. Following the notation of Ref. \citep{Tanay16}, we use the parameter $v$, defined by $v=(2\pi MF)^{1/3}=(2\pi M f/\ell)^{1/3}$, to refer to PN order. The constant $v_0$ is its initial value $(2\pi MF_0)^{1/3}$. On the other hand, the coefficients of PN parameter get frequency dependence at high order of eccentricity $(\geq 4)$. We use $\chi_\ell$ to represent such relationship. Both $v$ and $\chi_\ell$ depend on $\ell$. We thus obtain
\begin{align}
\Psi_\ell=&-2\pi ft_c+\ell l_c+ \Psi^{\GR}_{\ell} + e_{0}^{2} \Psi^\GRtwo_{\ell} + e_{0}^{4} \Psi^\GRfour_{\ell}+ e_{0}^{6} \Psi^\GRsix_{\ell} + e_{0}^{8} \Psi^\GReight_{\ell} + \Psi^{\BD}_{\ell}\,, 
\end{align}
where
\begin{align}
&\Psi^\GR_{\ell}=-\frac{\ell}{2}\frac{3}{128\eta v^5}\left[1+\frac{20}{9}\left(\frac{743}{336}+\frac{11}{4}\eta\right)v^2-16\pi v^3\right.+10\left(\frac{3058673}{1016064}+\frac{5429}{1008}\eta+\frac{617}{144}\eta^2\right)v^4+\pi\left(\frac{38645}{756}-\frac{65}{9}\eta\right) \notag \\
&\times\left\{1+3\log\left(\frac{v}{v_\text{lso}}\right)\right\}v^5+\left\{\frac{11583231236531}{4694215680}-\frac{640}{3}\pi^2\right.-\frac{6848}{21}\gamma-\frac{6848}{21}\log(4v)+\left(-\frac{15737765635}{3048192}+\frac{2255}{12}\pi^2\right)\eta \notag \\
&\left.\left.+\frac{76055}{1728}\eta^2-\frac{127825}{1296}\eta^3\right\}v^6\right],
\end{align}
\begin{align}
&\Psi^{\GRtwo}_{\ell} =\frac{\ell}{2}\frac{7065}{187136}\frac{1}{\eta v^5}\left(\frac{v_0}{v}\right)^{19/3}\left\{1+\left(\frac{299076223}{81976608}\right.+\frac{18766963}{2927736}\eta\right)v^2+\left(\frac{2833}{1008}-\frac{197}{36}\eta\right)v_0^2-\frac{2819123}{282600}\pi v^3+\frac{377}{72}\pi v_0^3 \notag \\
&+\left(\frac{16237683263}{3330429696}+\frac{24133060753}{971375328}\eta.+\frac{1562608261}{69383952}\eta^2\right)v^4+\left(\frac{847282939759}{82632420864}-\frac{718901219}{368894736}\eta-\frac{3697091711}{105398496}\eta^2\right)v^2v_0^2 \notag \\
&+\left(-\frac{1193251}{3048192}-\frac{66317}{9072}\eta+\frac{18155}{1296}\eta^2\right)v_0^4-\left(\frac{2831492681}{118395270}+\frac{11552066831}{270617760}\eta\right)\pi v^5+\left(-\frac{7986575459}{284860800}+\frac{555367231}{10173600}\eta\right)\pi v^3v_0^2 \notag \\
&+\left(\frac{112751736071}{5902315776}+\frac{7075145051}{210796992}\eta\right)\pi v^2v_0^3+\left(\frac{764881}{90720}-\frac{949457}{22680}\eta\right)\pi v_0^5+\left[-\frac{43603153867072577087}{132658535116800000}+\frac{536803271}{19782000}\gamma \right.\notag \\
&+\frac{15722503703}{325555200}\pi^2+\left(\frac{299172861614477}{689135247360}-\frac{15075413}{1446912}\pi^2\right)\eta+\frac{3455209264991}{41019955200}\eta^2+\frac{50612671711}{878999040}\eta^3+\frac{3843505163}{59346000 }\ln2 \notag \\
&-\frac{1121397129}{17584000}\ln3\left.+\frac{536803271}{39564000}\ln(16v^2)\right]v^6+\left(\frac{46001356684079}{3357073133568}+\frac{253471410141755}{5874877983744}\eta-\frac{1693852244423}{23313007872 }\eta^2 \right.\notag \\
&\left.-\frac{307833827417}{2497822272}\eta^3\right)v^4v_0^2-\frac{1062809371}{20347200}\pi^2v^3v_0^3+\left(-\frac{356873002170973}{249880440692736}-\frac{260399751935005}{8924301453312}\eta+\frac{150484695827}{35413894656}\eta^2\right. \notag \\
&\left.+\frac{340714213265}{3794345856}\eta^3\right)v^2v_0^4+\left[\frac{26531900578691}{168991764480}-\frac{3317}{126}\gamma+\frac{122833}{10368}\pi^2+\left(\frac{9155185261}{548674560}\right.-\frac{3977}{1152}\pi^2\right)\eta-\frac{5732473}{1306368}\eta^2-\frac{3090307}{139968}\eta^3 \notag \\
&+\frac{87419}{1890}\ln2\left.\left.-\frac{26001}{560}\ln3-\frac{3317}{252}\ln(16v_0^2)\right]v_0^6\right\},
\end{align}
\begin{align}
&\Psi^{\GRfour}_{\ell} =-\frac{\ell}{2}\frac{3}{128\eta v^5}\left\{\left(-\frac{2608555}{444448}\chi_\ell^{-19/9}+\frac{5222765}{998944}\chi_\ell^{-38/9}\right)\right.+\left[\left(-\frac{6797744795}{317463552}-\frac{426556895}{11337984}\eta\right)\chi_\ell^{-19/9} \right.+\left(-\frac{14275935425}{416003328} \right. \notag \\
&\left.+\frac{209699405}{4000032}\eta\right)\chi_\ell^{-25/9}+\left(\frac{198510270125}{10484877312}+\frac{1222893635}{28804608}\eta\right)\chi_\ell^{-38/9}\left.+\left(\frac{14796093245}{503467776}-\frac{1028884705}{17980992}\eta\right)\chi_\ell^{-44/9}\right]v^2 \notag \\
&+\left(\frac{217859203\pi}{3720960}\chi_\ell^{-19/9}-\frac{3048212305\pi}{64000512}\chi_\ell^{-28/9}-\frac{6211173025\pi}{102085632}\chi_\ell^{-38/9}+\frac{1968982405\pi}{35961984}\chi_\ell^{-47/9}\right)v^3+\left[\left(-\frac{369068546395}{12897460224}\right.\right. \notag \\
&\left.-\frac{548523672245}{3761759232}\eta-\frac{35516739065}{268697088}\eta^2\right)\chi_\ell^{-19/9}+\left(-\frac{37202269351825}{297145884672}-\frac{2132955527705}{74286471168}\eta+\frac{34290527545}{102041856}\eta^2\right)\chi_\ell^{-25/9} \notag \\
&+\left(-\frac{94372278903235}{7251965779968}+\frac{126823556396665}{733829870592}\eta-\frac{20940952805}{93768192}\eta^2\right)\chi_\ell^{-31/9}+\left(\frac{418677831611033}{34573325230080}+\frac{2163514670909}{12862100160}\eta \right. \notag \\
&\left.+\frac{203366083643}{1130734080}\eta^2\right)\chi_\ell^{-38/9}+\left(\frac{562379595264125}{5284378165248}+\frac{2965713234395}{94363895808}\eta-\frac{240910046095}{518482944}\eta^2\right)\chi_\ell^{-44/9}+\left(\frac{3654447011975}{98224939008}\right. \notag \\
&-\frac{4300262795285}{18124839936}\eta\left.\left.\left.+\frac{392328884035}{1294631424}\eta^2\right)\chi_\ell^{-50/9}\right]v^4\right\},
\end{align}
\begin{align}
&\Psi^{\GRsix}_{\ell} =-\frac{\ell}{2}\frac{3}{128\eta v^5}\left\{-\frac{1326481225}{101334144}\chi_\ell^{-19/9}+\frac{17355248095}{455518464}\chi_\ell^{-38/9}\right.-\frac{75356125}{3326976}\chi_\ell^{-19/3}+\left[\left(-\frac{3456734032025}{72381689856}-\frac{216909251525}{2585060352}\eta\right)\right. \notag \\
&\times\chi_\ell^{-19/9}+\left(-\frac{2441897241139735}{21246121967616}+\frac{9479155594325}{58368466944}\eta\right)\chi_\ell^{-25/9}+\left(\frac{659649627625375}{4781104054272}+\frac{4063675549105}{13134901248}\eta\right)\chi_\ell^{-38/9} \notag \\
&+\left(\frac{1968906345873305}{5969113952256}-\frac{8999675405695}{16398664704}\eta\right)\chi_\ell^{-44/9}+\left(-\frac{144936872901}{1691582464}-\frac{7378552295}{32530432}\eta\right)\chi_\ell^{-19/3}+\left(-\frac{213483902125}{1117863936}\right. \notag \\
&\left.+\left.\frac{14845156625}{39923712}\eta\right)\chi_\ell^{-7}\right]v^2+\left(\frac{22156798877\pi}{169675776}\chi_\ell^{-19/9}\right.-\frac{126468066221755\pi}{846342770688}\chi_\ell^{-28/9}-\frac{20639727962075\pi}{46551048192}\chi_\ell^{-38/9} \notag \\
&+\frac{33366234820475\pi}{65594658816}\chi_\ell^{-47/9}+\frac{30628811474315\pi}{97254162432}\chi_\ell^{-19/3}\left.-\frac{28409259125\pi}{79847424}\chi_\ell^{-22/3}\right)v^3+\left[\left(-\frac{187675742904025}{2940620931072}\right.\right. \notag \\
&\left.-\frac{278930807554775}{857681104896}\eta-\frac{18060683996675}{61262936064}\eta^2\right)\chi_\ell^{-19/9}+\left(-\frac{6363444229039638215}{15175834621968384}-\frac{39088433492776445}{270997046820864}\eta\right. \notag \\
&\left.+\frac{1550053258427425}{1488994762752}\eta^2\right)\chi_\ell^{-25/9}+\left(-\frac{387035983120116605285}{5846592827536441344}+\frac{1095104635088909345}{1338505683959808}\eta-\frac{185468261986684025}{191215097708544}\eta^2\right)\chi_\ell^{-31/9} \notag \\
&+\left(\frac{1391266434443462659}{15765436304916480}+\frac{7189359251430607}{5865117672960}\eta+\frac{675785495945689}{515614740480}\eta^2\right)\chi_\ell^{-38/9}+\left(\frac{74835480932061169625}{62651587527180288} \right. \notag \\
&\left.+\frac{14868442349448515}{21514968244224}\eta-\frac{2107245064767505}{472856444928}\eta^2\right)\chi_\ell^{-44/9}+\left(\frac{43949506831840859555}{63177102070677504}-\frac{1344731894414361455}{376054178992128}\eta\right. \notag \\
&\left.+\frac{7946157848161165}{2066231752704}\eta^2\right)\chi_\ell^{-50/9}+\left(-\frac{984783138418096685}{40879050017734656}-\frac{258954290041765}{271268315136}\eta-\frac{173415564792655}{148551696384}\eta^2\right)\chi_\ell^{-19/3} \notag \\
&+\left(-\frac{136868720309511}{189457235968}-\frac{17969188685519}{35523231744}\eta+\frac{1453574802115}{390365184}\eta^2\right)\chi_\ell^{-7}+\left(-\frac{26945014260125}{52819070976}+\frac{17350371000625}{6707183616}\eta\right.\notag \\
&\left.\left.\left.-\frac{357715525375}{119771136}\eta^2\right)\chi_\ell^{-23/3}\right]v^4\right\},
\end{align}
\begin{align}
&\Psi^{\GReight}_{\ell}=-\frac{\ell}{2}\frac{3}{128\eta v^5}\left(-\frac{250408403375}{1011400704}\chi_\ell^{-19/3}\right.+\frac{4537813337273}{39444627456}\chi_\ell^{-76/9}-\frac{6505217202575}{277250217984}\chi_\ell^{-19/9}\left.+\frac{128274289063885}{830865678336}\chi_\ell^{-38/9}\right).
\end{align}

\section{The temporal evolution of the orbital frequency and the eccentricity in MH coordinates}
\label{sec:Fdot-edot-MH}

We here present the coefficients that control the evolution of the orbital frequency and the eccentricity in Eqs.~\eqref{fdot-taylort4-inst}-\eqref{edot-taylort4-hered}. The latter equations had been presented in ADM coordinates before~\cite{Arun09}, but here we present them in MH coordinates.

\begin{align}
&\mathcal{O}_\text{N}=\frac{37e_t^4+292e_t^2+96}{5 \left(1-e_t^2\right)^{7/2}}, \\
&\mathcal{O}_\text{1PN}=\frac{1}{(1-e_t^2)^{9/2}}\left[-\left(\frac{1486}{35}+\frac{264}{5}\eta\right)+\left(\frac{2193}{7}-570\eta\right)e_t^2+\left(\frac{12217}{20}-\frac{5061}{10}\eta\right)e_t^4+\left(\frac{11717}{280} -\frac{148}{5}\eta\right)e_t^6\right], \\
&\mathcal{O}_\text{2PN}=\frac{1}{(1-e_t^2)^{11/2}}\left\{\left(-\frac{11257}{945}+\frac{15677}{105}\eta+\frac{944}{15}\eta^2\right)-\left(\frac{580291}{189}-\frac{2557}{5}\eta-\frac{182387}{90}\eta^2\right)e_t^2+\left(\frac{32657}{1260}-\frac{959279}{140}\eta\right.\right. \notag \\
&\left.+\frac{396443}{72}\eta^2\right)e_t^4+\left(\frac{4634689}{1680}-\frac{977051}{240}\eta+\frac{192943}{90}\eta^2\right)e_t^6+\left(\frac{391457}{3360}-\frac{6037}{56}\eta+\frac{2923}{45}\eta ^2\right)e_t^8+\left[\left(48-\frac{96}{5}\eta\right)\right. \notag \\
&\left.\left.+\left(2134-\frac{4268}{5}\eta\right)e_t^2+\left(2193-\frac{4386}{5}\eta\right)e_t^4+\left(\frac{175}{2}-35\eta\right)e_t^6\right]\sqrt{1-e_t^2}\right\}, \\
&\mathcal{O}_\text{3PN}=\frac{1}{(1-e_t^2)^{13/2}}\left\{\left[\frac{614389219}{148500}+\left(\frac{369}{2}\pi^2-\frac{57265081}{11340}\right)\eta-\frac{16073 \eta ^2}{140}-\frac{1121}{27}\eta^3\right]+\left[\frac{19898670811}{693000}+\left(\frac{3239}{16} \pi ^2\right.\right.\right. \notag \\
   &\left.\left.+\frac{2678401319}{113400}\right)\eta-\frac{9657701}{840}\eta ^2-\frac{1287385}{324}\eta ^3\right]e_t^2+\left[\frac{8036811073}{8316000}+\left(\frac{43741211273}{453600}-\frac{197087}{320} \pi ^2\right) \eta+\frac{1306589}{672}\eta ^2\right. \notag \\
   &\left.-\frac{33769597 \eta ^3}{1296}\right]e_t^4+\left[\frac{985878037}{5544000}+\left(\frac{54136669}{14400}-\frac{261211}{640}\pi^2\right)\eta+\frac{62368205\eta^2}{1344}-\frac{3200965 \eta ^3}{108}\right]e_t^6+\left[\frac{2814019181}{352000}\right. \notag \\
   &\left.-\left(\frac{12177}{640} \pi ^2+\frac{4342403}{336}\right)\eta+\frac{3542389 \eta ^2}{280}-\frac{982645 \eta ^3}{162}\right]e_t^8+\left(\frac{33332681}{197120}-\frac{1874543}{10080}\eta+\frac{109733}{840}\eta ^2-\frac{8288}{81} \eta ^3\right)e_t^{10} \notag \\
&+\sqrt{1-e_t^2}\left[-\frac{1425319}{1125}+\left(\frac{9874}{105} -\frac{41}{10}\pi ^2\right) \eta+\frac{632}{5}\eta^2+\left(\frac{933454}{375}+\frac{45961}{240}\pi^2\eta-\frac{2257181}{63}\eta+\frac{125278}{15}\eta ^2\right)e_t^2 \right.\notag \\
&+\left(\frac{840635951}{21000}-\frac{4927789}{60}\eta+\frac{6191}{32} \pi ^2 \eta+\frac{317273}{15}\eta^2\right)e_t^4 +\left(\frac{702667207}{31500}-\frac{6830419}{252}\eta+\frac{287}{960}\pi ^2 \eta+\frac{232177}{30}\eta ^2\right)e_t^6\notag \\
 &+\left.\left(\frac{56403}{112}-\frac{427733}{840}\eta+\frac{4739}{30}\eta ^2 \right)e_t^8 \right]+ \log
   \left[\frac{\sqrt{1-e_t^2}+1}{2 \left(1-e_t^2\right)}   \left(\frac{F(t)}{F_0}\right)^{2/3}\right]\left(\frac{54784}{175}+\frac{465664}{105}e_t^2+\frac{4426376}{525}
e_t^4 \right.\notag \\
&\left.\left.+\frac{1498856}{525}e_t^6+\frac{31779}{350}e_t^8\right)\right\}.
\end{align}
\begin{align}
&\mathcal{E}_\text{N}=\frac{1}{(1-e_t^2)^{5/2}}\left(\frac{304}{15}+\frac{121}{15}e_t^2\right), \\
&\mathcal{E}_\text{1PN}=\frac{1}{(1-e_t^2)^{7/2}}\left[-\left(\frac{939}{35}+\frac{4084}{45}\eta\right)+\left(\frac{29917}{105}-\frac{7753}{30}\eta\right)e_t^2+\left(\frac{13929}{280}-\frac{1664}{45}\eta\right)e_t^4\right], \\
&\mathcal{E}_\text{2PN}=\frac{1}{(1-e_t^2)^{9/2}}\left\{\left(-\frac{949877}{1890}+\frac{18763}{42}\eta+\frac{752}{5}\eta ^2\right)-\left(\frac{3082783}{2520}+\frac{988423}{840}\eta-\frac{64433}{40}\eta^2\right)e_t^2+\left(\frac{23289859}{15120}-\frac{13018711}{5040}\eta\right. \right. \notag \\
&\left.\left.\left.+\frac{127411}{90}\eta ^2\right)e_t^4+\left(\frac{420727}{3360}-\frac{362071}{2520}\eta +\frac{821}{9}\eta ^2\right)e_t^6 +\sqrt{1-e_t^2}\left[\frac{1336}{3}-\frac{2672}{15}\eta\right.+\left(\frac{2321}{2}-\frac{2321}{5}\eta\right)e_t^2+\left(\frac{565}{6}-\frac{113}{3}\eta\right)e_t^4 \right]\right\}, \notag \\
\end{align}
\begin{align}
&\mathcal{E}_\text{3PN}=\frac{1}{(1-e_t^2)^{11/2}}\left\{\frac{48189239083}{6237000}+\left(\frac{4469 \pi ^2 }{36}-\frac{266095577}{113400}\right)\eta-\frac{1046329 }{2520}\eta ^2-\frac{61001 }{486}\eta ^3+\left[\frac{24647957401}{1247400}+\left(\frac{475211069}{18900}\right.\right.\right.\notag \\
&\left.\left.-\frac{26773 \pi ^2}{288} \right)\eta-\frac{2020187 \eta ^2 }{3024}-\frac{86910509 \eta ^3}{19440}\right]e_t^2+\left[-\frac{71647254149}{16632000}+\left(\frac{3699068059}{907200}-\frac{1375673 \pi ^2}{5760}\right)\eta+\frac{253550327 \eta ^2}{20160}\right. \notag \\
&\left.-\frac{2223241 \eta ^3}{180}\right]e_t^4+\left[-\frac{41527976119}{4752000}+\left(\frac{120327659}{15120}-\frac{63673 \pi ^2}{5760}\right)\eta+\frac{12599311 \eta ^2}{2520} -\frac{11792069 \eta ^3}{2430}\right]e_t^6+\left(-\frac{684831911}{1774080}\right.\notag \\
&\left.+\frac{1290811 \eta}{3360}+\frac{2815 \eta ^2}{216}-\frac{193396 \eta ^3}{1215 }\right)e_t^8+\frac{\sqrt{1-e_t^2}}{\sqrt{1-e_t^2}+1}\left[-\frac{192764352 }{6237000}+\left(-\frac{2614513}{2625}+\frac{8323 \pi ^2 \eta}{180}-\frac{526991 \eta}{189}\right.\right.\notag \\
   &\left.+\frac{54332 \eta ^2 }{45}\right)e_t^2 +\left(\frac{89395687}{7875}+\frac{94177 \pi ^2 \eta}{960}-\frac{1871861 \eta}{210}+\frac{681989 \eta ^2 }{90}\right)e_t^4+\left(\frac{5321445613}{378000}+\frac{2501 \pi ^2 \eta }{2880}-\frac{26478311 \eta}{1512}\right.\notag \\
   &\left.\left.+\frac{225106 \eta ^2 }{45}\right)e_t^6+\left(\frac{186961}{336}-\frac{289691 \eta}{504}+\frac{3197 \eta ^2 }{18}\right)e_t^8\right]+\log
   \left[\left(\frac{F}{F_0}\right)^{2/3}\frac{\sqrt{1-e_t^2}+1}{2
   \left(1-e_t^2\right)}\right]\left(\frac{1316528}{1575}+\frac{4762784}{1575}e_t^2\right.\notag \\
   &\left.\left.+\frac{2294294 }{1575}e_t^4+\frac{20437 }{350}e_t^6 \right)\right\}.
\end{align}
\end{widetext}

\section{The dependence of the overlap on the mass ratio}
\label{sec:app-overlap}
\begin{figure}[!h]
\centering
\includegraphics[width=\columnwidth,clip=true]{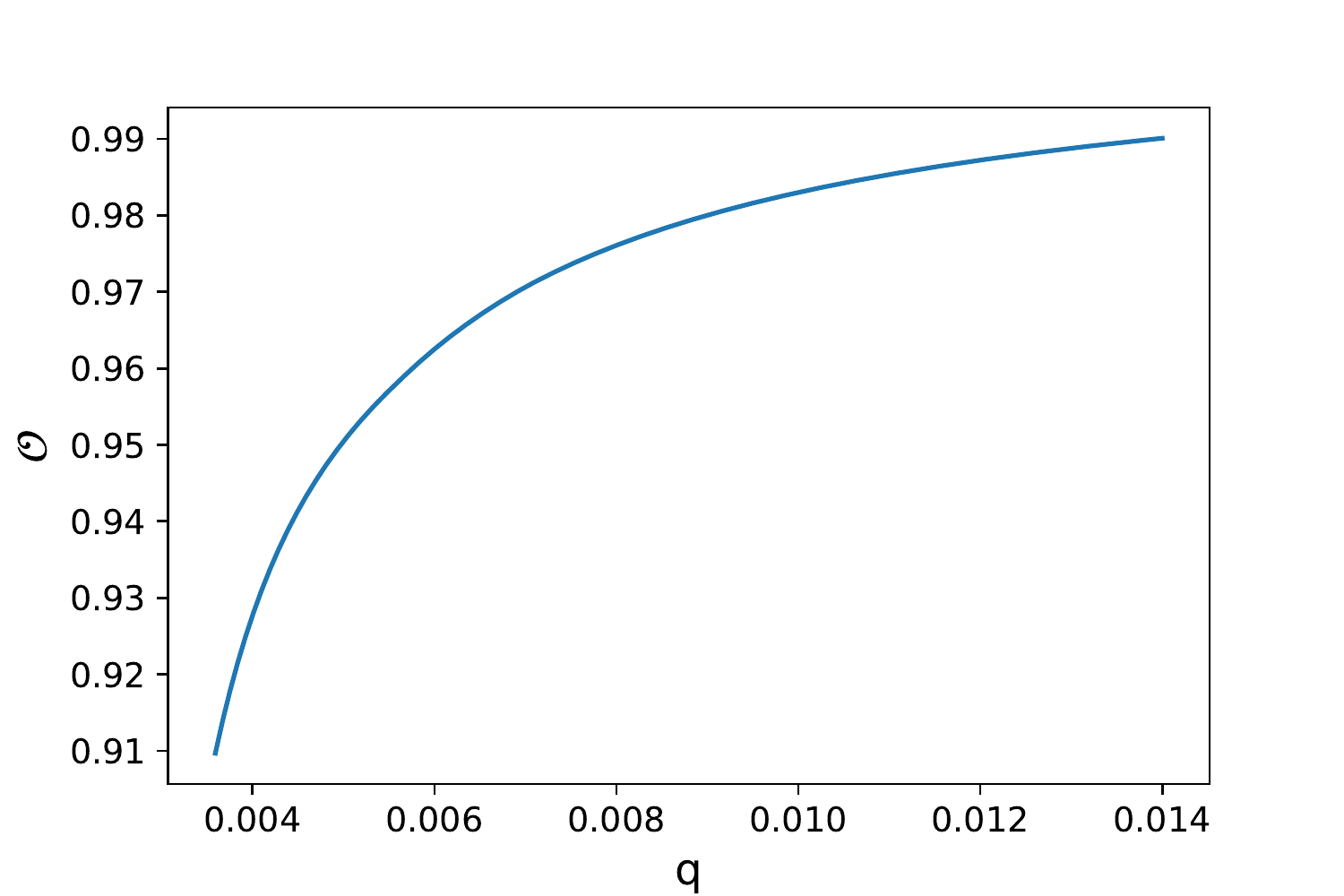}
\caption{(Color Online) The match between a 3PN TalyorT4 and a 3PN TaylorF2 model for quasi-circular inspirals as a function of the mass ratio $q$.}
\label{fig:q-overlap}
\end{figure}

In this appendix, we discuss how the overlap changes with mass ratio, focusing on GWs from quasi-circular binaries. We consider 5-year long LISA signals generated by BHNS binary inspirals, with the NS mass fixed at $m_2=1.4M_\odot$. We vary the BH mass $m_1$ and plot the overlap $\mathcal{O}$ as a function of the mass ratio $q=m_2/m_1$, as shown in Fig. \ref{fig:q-overlap}. Observe that the overlap increases monotonically with mass ratio $q$. When $q=7.0\times10^{-3}$, the overlap equals the 0.97 threshold. This indicates that if the mass ratio is small enough, the overlap between the TaylorF2 and TaylorT4 models becomes sufficiently small that the analytic model need not be sufficiently accurate any longer. The breakdown of the PN approximation for small mass ratios $q$ is known in the EMRI literature and it should be addressed elsewhere.
\FloatBarrier


\bibliography{refer}
\bibliographystyle{apsrev}
\end{document}